\documentclass[nofootinbib,aps,pra,superscriptaddress,reprint]{revtex4-1}
\usepackage{amsmath,amsthm}
\usepackage{graphicx}
\usepackage{color}
\usepackage{amsfonts,amssymb}
\usepackage{bm}
\usepackage{braket}
\usepackage{mathptmx}
\usepackage{fancyhdr} 
\usepackage{soul} 
\usepackage{times}

\usepackage[colorlinks,linkcolor=blue,anchorcolor=blue,citecolor=blue,urlcolor=black]{hyperref}

\begin{document}

\title{Noiseless Linear Amplifiers for Multimode States}

\date{\today}

\author{Mingjian He}
\email{mingjian.he@unsw.edu.au}

\author{Robert Malaney}
\email{r.malaney@unsw.edu.au}
\affiliation{School of Electrical Engineering and Telecommunications,\\
			The University of New South Wales, Sydney, NSW 2052, Australia.}

\author{Benjamin A. Burnett}
\email{benjamin.burnett@ngc.com}
\affiliation{Northrop Grumman Corporation, San Diego, CA 92127, USA.}

\begin{abstract}
The entanglement structure between different frequency components within broadband quantum light pulses, forged at entanglement creation, represents a promising route to the practical delivery of many multipartite quantum information applications.
However, the scalability of such applications is largely limited by the entanglement decoherence caused by photon loss.
One promising method to combat such losses is noiseless linear amplification.
However, while there have been various procedures that implement noiseless linear amplification on single-mode states, no realization has thus far been proposed for noiseless linear amplification on quantum states carrying a multimode structure. In this work we close this gap, proposing a novel Noiseless Linear Amplifier (NLA) with Photon Catalysis (PC), namely, the PC-NLA.
Constructing a multimode version of an existing NLA that uses Quantum Scissors (QS), the QS-NLA, we then show how the PC-NLA is compatible with the QS-NLA, even though the former uses half the physical resources of the latter.
We then apply our newly developed multimode NLA frameworks to the problem of Continuous-Variable (CV) entanglement distillation, determining how the multimode structure of the entanglement impacts the performance of the NLAs. Different from single-mode NLA analyses,
we find that a multimode NLA is only effective as a CV entanglement distillation strategy when the channel loss is beyond some threshold - a threshold largely dependent on the multimode structure.
The results provided here will be valuable for real-world implementations of multipartite quantum information applications that utilize complex entanglement structure within broadband light pulses.

\end{abstract}

\maketitle

\thispagestyle{fancy}
\renewcommand{\headrulewidth}{0pt}

\section{Introduction}
In general, a quantum state encapsulated within a light pulse consists of multiple frequency modes (a multimode) - a reality of special importance for a plethora of Continuous Variable (CV) quantum information applications that utilize high-pulse-rate laser sources. For example, entanglement between different \emph{supermodes} (linear combinations of frequency modes) of ultra-fast light pulses could allow for a practical route to higher quantum information throughput \cite{christ2012exponentially,hosseinidehaj2017multimode,kumar2019continuous}, be used as a resource for quantum computing \cite{menicucci2008one,armstrong2012programmable,ferrini2013compact,chen2014experimental}, enable improved quantum sensing \cite{zhuang2018distributed,xia2019repeater,guo2020distributed,gessner2020multiparameter}, and be used in quantum secret sharing \cite{cai2017multimode}.
State-of-the art laser sources are now in the 100MHz regime with pulse rates of 1GHz, and beyond, anticipated as mainstream in the coming {years} \cite{fortier201920}. In addition to random entanglement between supermodes, the broadband nature of ultra-fast light pulses represents a fertile ground for developing precisely-engineered CV entanglement across frequency space \cite{de2014full}.
Parametric-down-conversion of ultra-fast frequency-combs provides one convenient single-step path to specifically engineered entangled-supermode states \cite{roslund2014wavelength,gerke2015full}.

Like any other entangled states, multimode states suffer from decoherence over lossy channels.
One route open to compensate such loss is amplification of the states.
Ideally, such amplifiers would be noiseless and linear (phase-insensitive) - a goal that is impossible in a deterministic sense \cite{caves1982quantum}.
However, it is known that a noiseless linear amplifier (NLA) acting on quantum states is  possible in a non-deterministic, or probabilistic fashion, e.g. \cite{ralph2009nondeterministic}.
What remains to be determined is how to construct a probabilistic NLA for multimode states, and what the performance of such a multimode NLA is relative to a single-mode (single frequency) NLA - especially with regard to important CV information protocols such as entanglement distillation.
This task forms the focus of this work.

In order to proceed towards a multimode NLA, we will consider two approaches  previously developed for single-mode systems - a Quantum Scissors (QS) approach \cite{ralph2009nondeterministic} and a Photon Catalysis (PC) approach \cite{zhang2018photon}. QS, a technique which operates using beam-splitters and single-photon detectors \cite{pegg1998optical}, is known to operate as an NLA under the limitation of low-energy input states \cite{ralph2009nondeterministic}.
PC, which requires  half the number of beam-splitters and single-photon detectors required by QS, is also known to act as an NLA for low-energy input states, but with higher probabilities of success relative to QS \cite{zhang2018photon}.\footnote{A cascaded application of PC was investigated in terms of entanglement distillation in \cite{mardani2020continuous}, but found to bring insignificant enhancement relative to the single-use of PC.}
In order to overcome the limitation of QS-based NLAs being applicable only to low-energy inputs, \cite{ralph2009nondeterministic} proposed the notion of parallel-processing QS.
Experimental realizations, at least in part, based on the ideas in \cite{ralph2009nondeterministic} have been undertaken, e.g. \cite{ferreyrol2010implementation,zavatta2011high,chrzanowski2014measurement,gagatsos2014heralded,ulanov2015undoing,haw2016surpassing}.
{Different schemes have also been proposed for implementing the NLA}, e.g. \cite{mivcuda2012noiseless,kim2012quantum,yang2013improving,zhao2017characterization,hu2019entanglement,winnel2020generalised}.
All of the above work, however, is designed with the single-mode picture in view. To analyse the more generic picture, extensions to a multimode framework are required.

The key contributions of this work can be summarized as follows:
We propose a novel NLA based on parallel processing of PC, the PC-NLA.
We show that when applied to a range of coherent states, the PC-NLA is compatible with the QS-NLA even though the former requires less physical resources than the latter. We then assess how the  multimode structure of states impacts the performance of the PC-NLA and QS-NLA when applied to entanglement distillation -- showing how {a multimode analysis} can lead to outcomes quite distinct from a single-mode {analysis}. A multimode framework for QS is developed in order to carry out this assessment.

The rest of this paper is organized as follows.
In Section~\ref{sec:multiNLA} we present the multimode version of the QS-NLA.
We then show {how} PC can be utilized to build the multimode PC-NLA.
In Section~\ref{sec:compare} we compare the performances of the QS-NLA and the PC-NLA with regard to entanglement enhancement.
In Section~\ref{sec:parcas} we compare our PC-NLA and a recently proposed cascaded processing of PC.
Section~\ref{sec:conclusion} concludes our work.

\section{Noiseless linear amplifiers for\\ CV quantum states}\label{sec:multiNLA}
\subsection{NLA with QS}\label{sec:multiqs}

\begin{figure}
	\centering
	\includegraphics[width=1\linewidth]{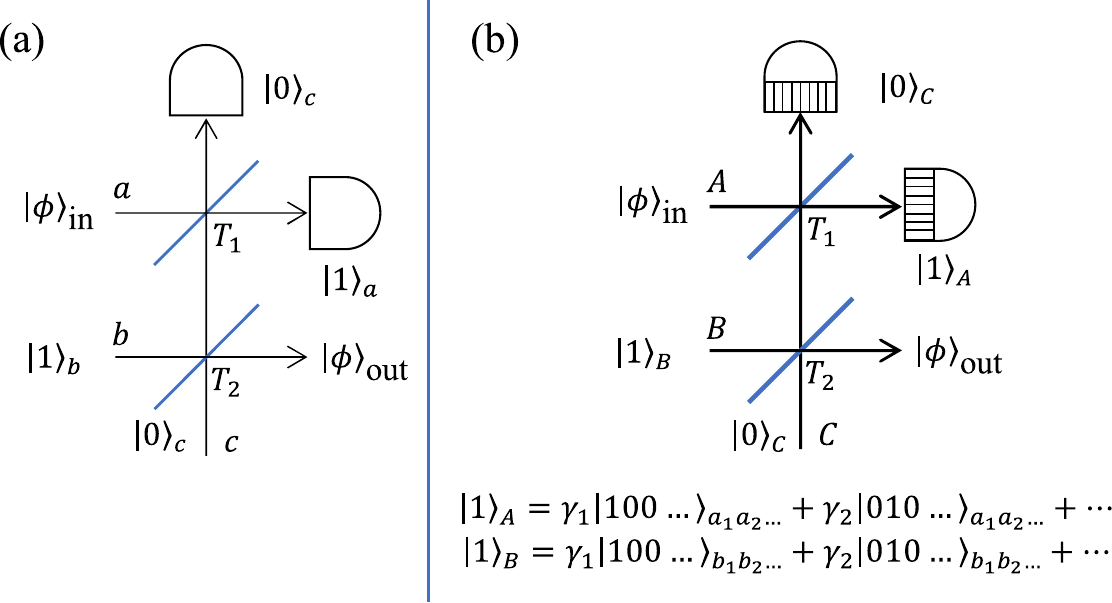}
	\caption{QS for (a) single-mode states and (b) multimode states.
		For the multimode case the single-photon detector {implements a joint detection on the single-mode components of the multimode state and clicks when a photon with a specific multimode structure (determined by the weighting coefficients $\gamma_1,\gamma_2,...$) is detected. An implementation of the multimode detector is discussed in section \ref{sec:compare}.}}
	\label{fig:diagqs}
\end{figure}

In this section, we propose an NLA with parallel QS for multimode states.
But first let us briefly review the well-studied QS operation for single-mode states (Fig.~\ref{fig:diagqs}a).

\textit{1) Single-mode case:}
Let $a$, $b$, and $c$ label the single-modes involved in the operation.
In the Fock basis, the QS can be represented by an operator \cite{hu2019entanglement}
\begin{equation}\label{eq:singleQS}
\begin{aligned}
\hat M=&\bra{0}_c\bra{1}_a
U_{ac}(T_1)
U_{bc}(T_2)
\ket{1}_b\ket{0}_c\\
=&\sqrt{T_1T_2}\ket{0}_b\bra{0}_a+\sqrt{(1-T_1)(1-T_2)}\ket{1}_b\bra{1}_a,
\end{aligned}
\end{equation}
where $U_{ac}(T_1)$ and $U_{bc}(T_2)$ are the unitary operators of the beam-splitters (shown in Fig.~\ref{fig:diagqs}a), and $T_1$ and $T_2$ are their transmissivities.
Conventionally, it is assumed that $T_1=1/2$, in which case $\hat{M}$ reduces to
\begin{equation}\label{eq:singleQST}
\begin{aligned}
\hat M=\sqrt{\frac{T}{2}}\ket{0}_b\bra{0}_a+\sqrt{\frac{1-T}{2}}\ket{1}_b\bra{1}_a.
\end{aligned}
\end{equation}
QS can also be implemented by detecting a single photon at output port $c$ and zero photons at output port $a$.
In this case, the operator for the QS can be written as
\begin{equation}
\begin{aligned}
\hat M'=\sqrt{\frac{T}{2}}\ket{0}_b\bra{0}_a-\sqrt{\frac{1-T}{2}}\ket{1}_b\bra{1}_a,
\end{aligned}
\end{equation}
which can be converted to the previous case (in Eq.~(\ref{eq:singleQST})) by adding a phase shifter at the output of the QS.
We note the conversion between the two cases is only possible under the setting of $T_1=1/2$.
The success probabilities for the two cases are identical.

\begin{figure}
	\centering
	\includegraphics[width=1\linewidth]{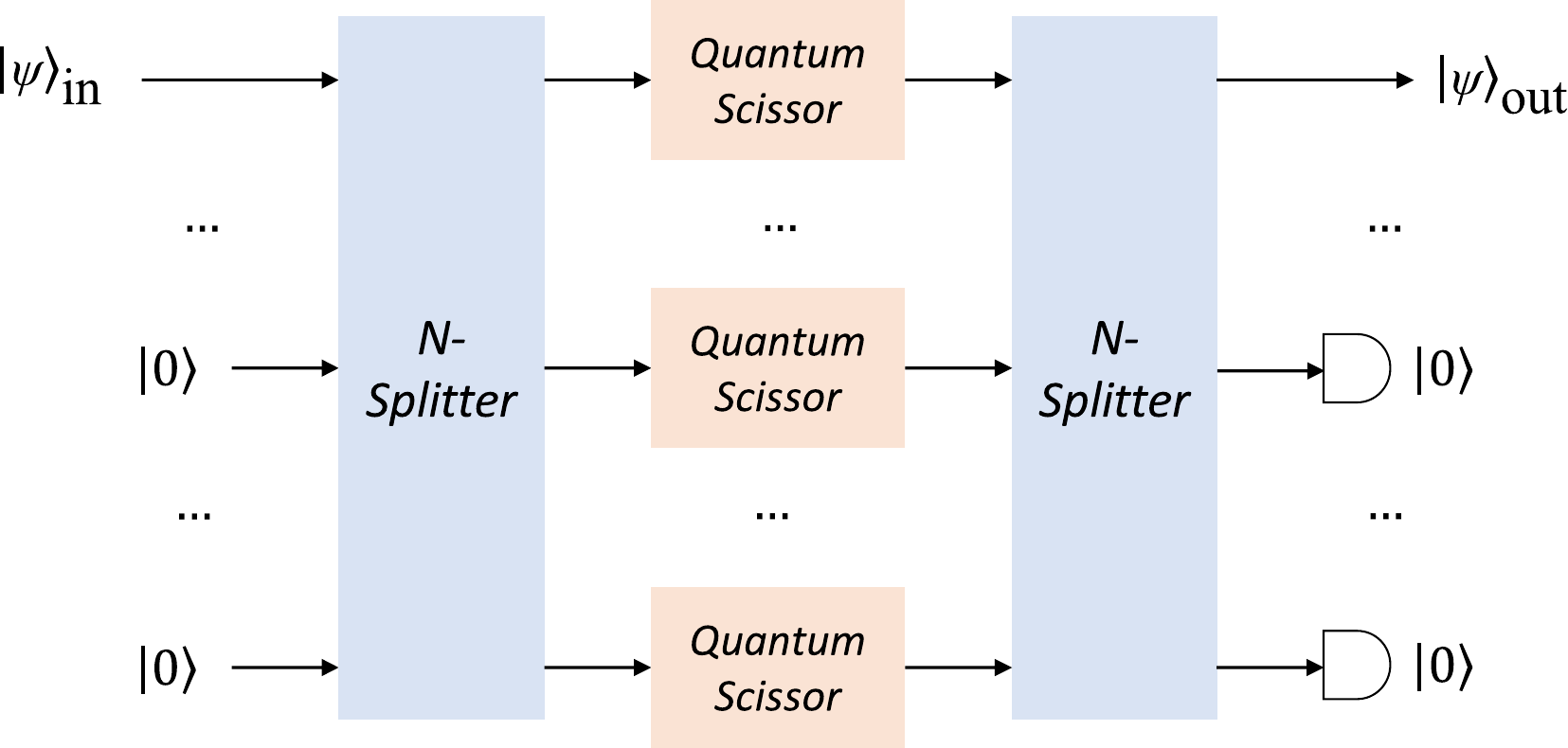}
	\caption{The QS-NLA.}
	\label{fig:diagnqs}
\end{figure}

A limitation for QS is that it can only operate as an NLA for weak states,\footnote{NLAs are normally considered with reference to amplification of a coherent state $\left| \alpha  \right\rangle$. In this context a weak state means $\left| \alpha  \right| <  < 1$.}
 i.e., $\ket{\psi}_{in}\approx\ket{0}+\alpha\ket{1}$, where $\alpha$ is a complex number.
A device, which we refer to as the QS-NLA, was proposed in \cite{ralph2009nondeterministic} to overcome this limitation.
As is depicted in Fig.~\ref{fig:diagnqs}, one major component of the QS-NLA is the $N$-splitter, which consists of an array of beam-splitters. The first $N$-splitter, in conjunction with the vacuum ancillae, evenly divides the input state into $N$ paths.
Parallel QS operations are then applied to each path.
{The transmissivities $T$ for the beam-splitters in the QS operations are identical.}
The second $N$-splitter adopts an inverse to the arrangement of the beam-splitters in the first $N$-splitter.
The paths after the QS operations are interferometrically recombined at the second $N$-splitter.
The amplification is successful when all the output ports except the first port ($\ket{\psi}_{\text{out}}$) of the second $N$-splitter register zero photons.
The QS-NLA approaches an ideal NLA when $N$ is large, but its success probability vanishes as $N$ grows.
For a coherent state input, the QS-NLA implements the following transformation (in the limit of $N>>\sqrt{\frac{1-T}{T}}|\alpha|$) \cite{ralph2009nondeterministic}
\begin{equation}\label{eq:qstransform}
\ket{\alpha}\rightarrow
\frac{1}{\sqrt{P}}\sqrt{T}^N e^{-\frac{\left(1-g_s^{2}\right)|\alpha|^{2}}{2}}\ket{g_s \alpha},
\end{equation}
which has the success probability
\begin{equation}\label{eq:qsprob}
P=T^{N} e^{-\left(1-g_s^{2}\right)|\alpha|^{2}},
\end{equation}
where the equivalent gain reads $g_s=\sqrt{(1-T)/T}$.

The QS-NLA can be directly extended to the multimode case by replacing the single-mode QS with the multimode QS as discussed next.

\textit{2) Multimode case:} In the terminology adopted in this work, a multimode  is simply a generic collection of single-modes.
A supermode  refers to a specific linear superposition of single-modes - with a sequence of such supermodes forming an orthonormal basis.
Let $\hat{a}^\dagger_m$ be the creation operator of a single-mode  at a specific frequency (indexed with $m\in\left\{1,2,...,\infty\right\}$)\footnote{{In experiments $m$ will be an index labeling a limited number of discrete frequency bins, the bandwidth being determined by the resolution of the detectors.}}, then a new creation operator can be defined as
\begin{equation}\label{eq:multicreation}
\hat{A}^\dagger=\sum_{m=1}^{\infty}\gamma_m a_m^\dagger,
\end{equation}
where the $\gamma_m$'s are normalized complex weighting coefficients satisfying $\sum_{m=1}^\infty|\gamma_m|^2=1$.
The composed mode created by $\hat{A}^\dagger$ is what we refer to as a supermode.

For the multimode QS, two major components appearing in the single-mode QS need to be generalized.
These are the Fock state and the beam-splitter operator.
We first define the multimode Fock state as
\begin{gather}\label{eq:mutliSingle}
\begin{aligned}
\ket{n}_A=\frac{\hat{A}^{\dagger n}}{\sqrt{n!}}&\ket{0}_A,
\end{aligned}
 \end{gather}
where
\begin{equation}
\ket{0}_A=\bigotimes_{m=1}^\infty \ket{0}_{a_m},
\end{equation}
and we have used capital letters as subscripts to label the supermodes.
Specifically, a multimode single-photon state can be represented by
\begin{equation}
\ket{1}_A=\sum_{m'=1}^{\infty}\gamma_{m'}\bigotimes_{m=1}^\infty \ket{\delta_{m,m'}}_{a_m},
\end{equation}
where $\delta_{m,m'}=1$ for $m=m'$, and $\delta_{m,m'}=0$ for $m\neq m'$.

{We assume the beam-splitter in the multimode QS operation is frequency independent such that the transmissivities associated with each single-mode component of the multimode state are identical.}
The multimode beam-splitter operator can then be written as
\begin{equation}\label{eq:multiBS}
\hat{\textbf{U}}_{AB}\left( {T} \right)=\bigotimes_{m=1}^\infty \hat{U}_{a_mb_m}\left( T \right),
\end{equation}
where
$\hat{U}_{a_mb_m}\left( T \right)$ is the single-mode beam-splitter operator coupling two single-modes $a_m$ and $b_m$.

As illustrated in Fig.~\ref{fig:diagqs}b, let $A$, $B$, and $C$ label the supermodes involved in the QS operation.
The multimode QS can then be represented by an operator (assuming $T_1=1/2$ and $T_2=T$).
\begin{equation}\label{eq:multiQS}
\begin{aligned}
\hat M=&\bra{0}_C\bra{1}_A
\textbf U_{AC}(T_1)
\textbf U_{BC}(T_2)
\ket{1}_B\ket{0}_C\\
=&\sqrt{\frac{T}{2}}\ket{0}_B\bra{0}_A+\sqrt{\frac{1-T}{2}}\ket{1}_B\bra{1}_A,
\end{aligned},
\end{equation}
where
\begin{equation}
\ket{1}_B=\hat{B}^\dagger\ket{0}_B=\sum_{m=1}^{\infty}\gamma_m b_m^\dagger\ket{0}_B.
\end{equation}
The multimode QS operator reduces to the single-mode QS operator in Eq.~(\ref{eq:singleQS}) in the special case where each supermode only has one single-mode component.
The derivation for Eq.~(\ref{eq:multiQS}) can be found in Appendix \ref{ap:1}.

Consider a multimode coherent state, which can be expressed in the Fock basis by
\begin{equation}\label{eq:multicoherent}
\begin{aligned}
\ket{\alpha}_A&=e^{-\frac{|\alpha|^2}{2}}\sum_{n=0}^{\infty} \frac{\alpha^n}{\sqrt{n!}}\ket{n}_A\\
&=e^{-\frac{|\alpha|^2}{2}}\sum_{n=0}^{\infty} \frac{\left(\alpha A^\dagger\right)^n}{n!}\ket{0}_A,
\end{aligned}
\end{equation}
where $A^\dagger$ is the supermode mode creation operator defined in Eq.~(\ref{eq:multicreation}).
Similar to the single-mode case, a multimode QS-NLA can implement the following transformation
\begin{equation}\label{eq:qstransformmulti}
\ket{\alpha}_A\rightarrow
\frac{1}{\sqrt{P}}\sqrt{T}^N
e^{-\frac{\left(1-g_s^{2}\right)|\alpha|^{2}}{2}}\ket{g_s \alpha}_A,
\end{equation}
with the same success probability as in Eq.~(\ref{eq:qsprob}), where again $g_s=\sqrt{(1-T)/T}$.

\subsection{NLA with PC}
We now propose a novel NLA with parallel PC operations.
Noticing that the single-mode case can be viewed as a special case in the multimode setting, we discuss this new amplifier in the multimode setting only.
In the rest of the paper, for brevity we will omit the subscripts for the supermodes.

\begin{figure}
	\centering
	\includegraphics[width=1\linewidth]{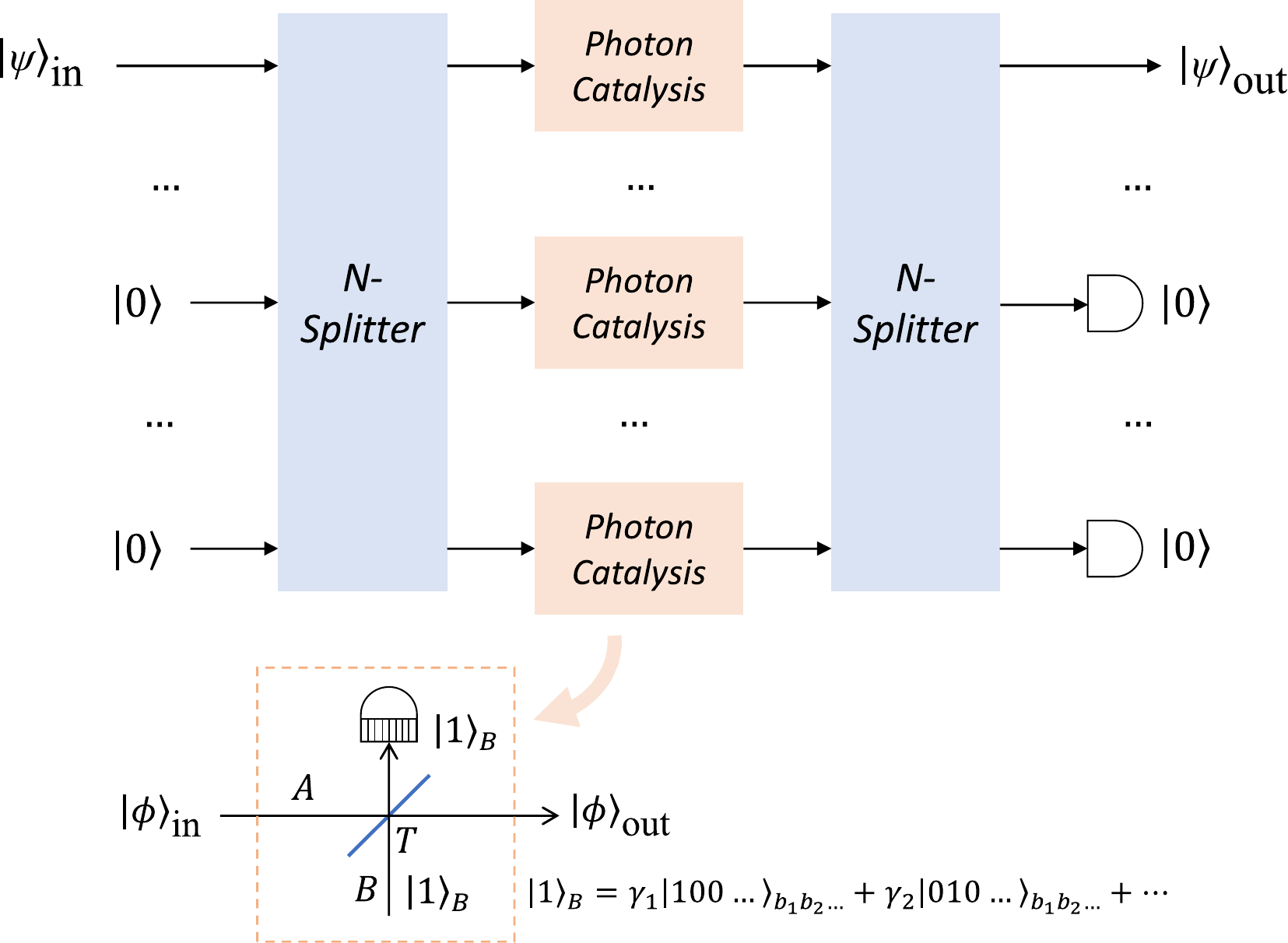}
	\caption{The diagrams for the PC-NLA.}
	\label{fig:diagncatalysis}
\end{figure}

An NLA can be constructed with PC.
We refer to this type of NLA as a PC-NLA, {a schematic of which is given in} Fig.~\ref{fig:diagncatalysis}.
An advantage of the PC-NLA is that each PC module only requires one beam-splitter and one single-photon detector, which is half of the apparatus required by {a QS module. This reduction in apparatus can be important, especially in confined environments (e.g. on-board a satellite).}

We now investigate a scenario where the state to be amplified is the multimode coherent state defined in Eq.~(\ref{eq:multicoherent}).
The first $N$-splitter of Fig.~\ref{fig:diagncatalysis} divides the input coherent state into the product state, that is
\begin{equation}
\ket{\alpha} \longrightarrow \ket{\alpha'}^{\otimes N},
\end{equation}
where $\alpha'=\alpha/\sqrt{N}$.
Parallel PC operations are then applied on each $\ket{\alpha'}$, where each PC operation can be represented by an operator \cite{he2020global}
\begin{equation}\label{eq:multiR}
\hat{R}=\sqrt{T}\left(-\frac{1-T}{T}\hat{A}^\dagger\hat{A}+1\right)\bigotimes_{m=1}^{\infty}\sqrt{T}^{
\hat{a}_m^\dagger\hat{a}_m},
\end{equation}
and where $T$ is the transmissivity for the beam-splitter in the PC operation.
{The rightmost term in the above equation satisfies the following identities}
\begin{equation}
\bigotimes_{m=1}^{\infty}\sqrt{T}^{\hat{a}_m^\dagger\hat{a}_m}\ket{n}
=\sqrt{T}^{\hat{A}^\dagger\hat{A}}\ket{n}=\sqrt{T}^n \ket{n}.
\end{equation}
{Similar to the QS-NLA, the transmissivities $T$ for the beam-splitters in the PC operations are identical.}
Each PC operation will alter each $\ket{\alpha'}$ to (not-normalized)
\begin{equation}
\hat{R}\ket{\alpha'}
=e^{-\frac{(1-T)|\alpha'|^2}{2}}
\sqrt{T}\left(-\frac{1-T}{T} \alpha'' \hat{A}^\dagger +1\right)\ket{\alpha''},
\end{equation}
where $\alpha''=\sqrt{T}\alpha/\sqrt{N}$.
The photon-catalyzed coherent state is then coherently recombined at the second $N$-splitter.
Post-selecting the state in the $\ket{\psi}_{\text{out}}$ port only when the other output ports register zero photons, leads to an output state
\begin{equation}
\begin{aligned}
\ket{\psi}_{out}
=&\frac{1}{\sqrt{P}}\sqrt{T}^N
e^{-\frac{|\alpha|^2}{2}}\\
&\times
\left(-\frac{1-T}{\sqrt{T}}\frac{\alpha}{N} \hat{A}^\dagger + 1 \right)^N
e^{\sqrt{T}\alpha \hat{A}^\dagger}
\ket{0},
\end{aligned}
\end{equation}
where $P$ is the success probability for the PC-NLA.
In the limit of $N>>\frac{1-T}{\sqrt{T}}|\alpha|$, we have
\begin{equation}
\begin{aligned}
\lim_{N>>\frac{1-T}{\sqrt{T}}|\alpha|} & \left(-\frac{1-T}{\sqrt{T}}\frac{\alpha}{N} \hat{A}^\dagger + 1 \right)^N
e^{\sqrt{T}\alpha \hat{A}^\dagger}
\ket{0}\\
&=e^{-\frac{1-T}{\sqrt{T}}{\alpha}\hat{A}^\dagger}
e^{\sqrt{T}\alpha \hat{A}^\dagger}
\ket{0}\\
&=e^{\frac{|g_c\alpha|^2}{2}}
\ket{-g_c \alpha},
\end{aligned}
\end{equation}
where the equivalent gain reads $g_c=(1-2T)/ \sqrt{T}$.
Putting everything together we conclude in the limit of $N>>\frac{1-T}{\sqrt{T}}|\alpha|$, the PC-NLA implements the transformation
\begin{equation}\label{eq:pctransformmulti}
\begin{aligned}
\ket{\alpha}\rightarrow & \frac{1}{\sqrt{P}}\sqrt{T}^N
e^{-\frac{(1-g_c^2)|\alpha|^2}{2}}
\ket{-g_c \alpha},
\end{aligned}
\end{equation}
where
\begin{equation}
P=T^N e^{-(1-g_c^2)|\alpha|^2}.
\end{equation}
The PC-NLA acts as an NLA up to an irrelevant global phase. The equivalent gain of the PC-NLA satisfies $g_c>1$ when $T<0.25$.

From Eq.~(\ref{eq:qstransformmulti}) and Eq.~(\ref{eq:pctransformmulti}) it can be observed that the QS-NLA and the PC-NLA implement similar transformations.
At fixed $T$, the only difference for the two NLAs is their equivalent gains ($g_s$ and $g_c$).
For an input coherent state, it can be shown that the QS-NLA always has a higher success probability than the PC-NLA when $g_s=g_c$.

\subsection{Comparison of the Two NLAs for Finite $N$}
\begin{figure*}
	\centering
	\includegraphics[width=.32\linewidth]{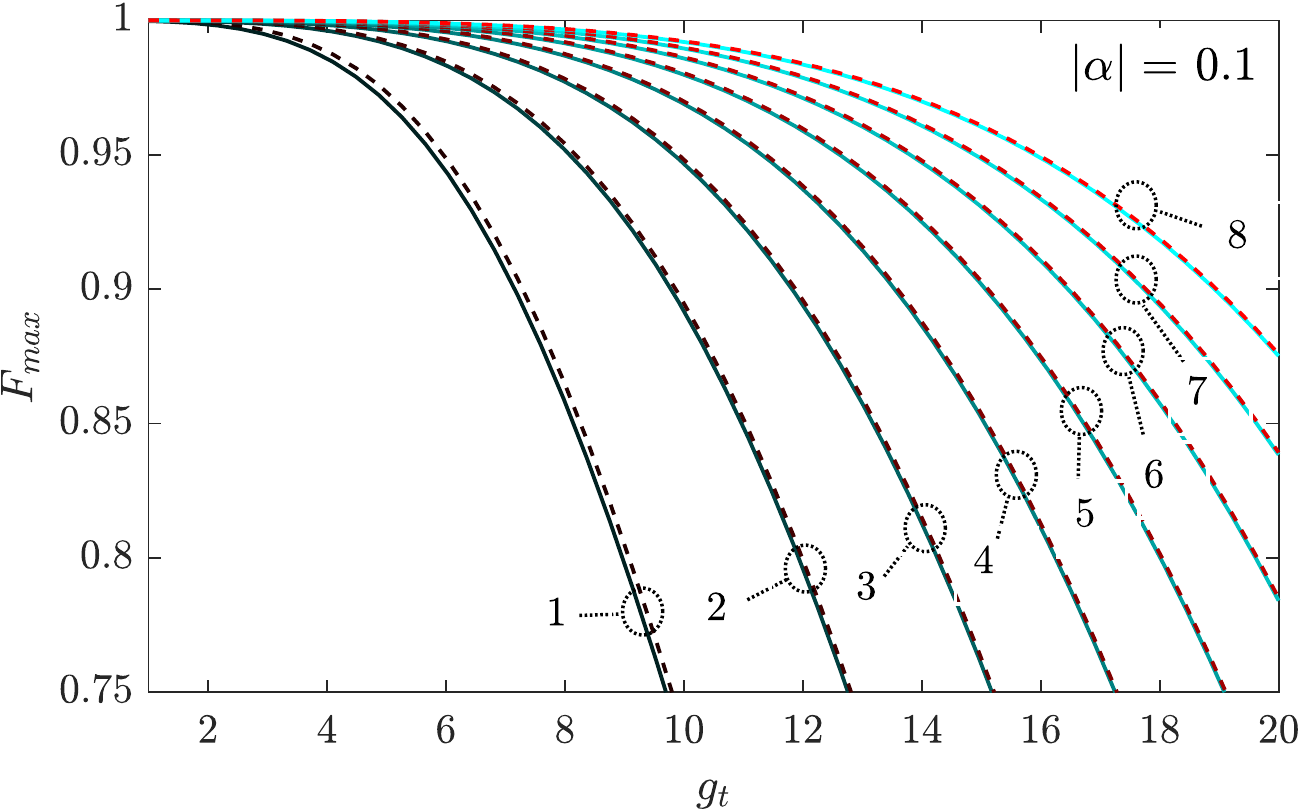}
	\includegraphics[width=.32\linewidth]{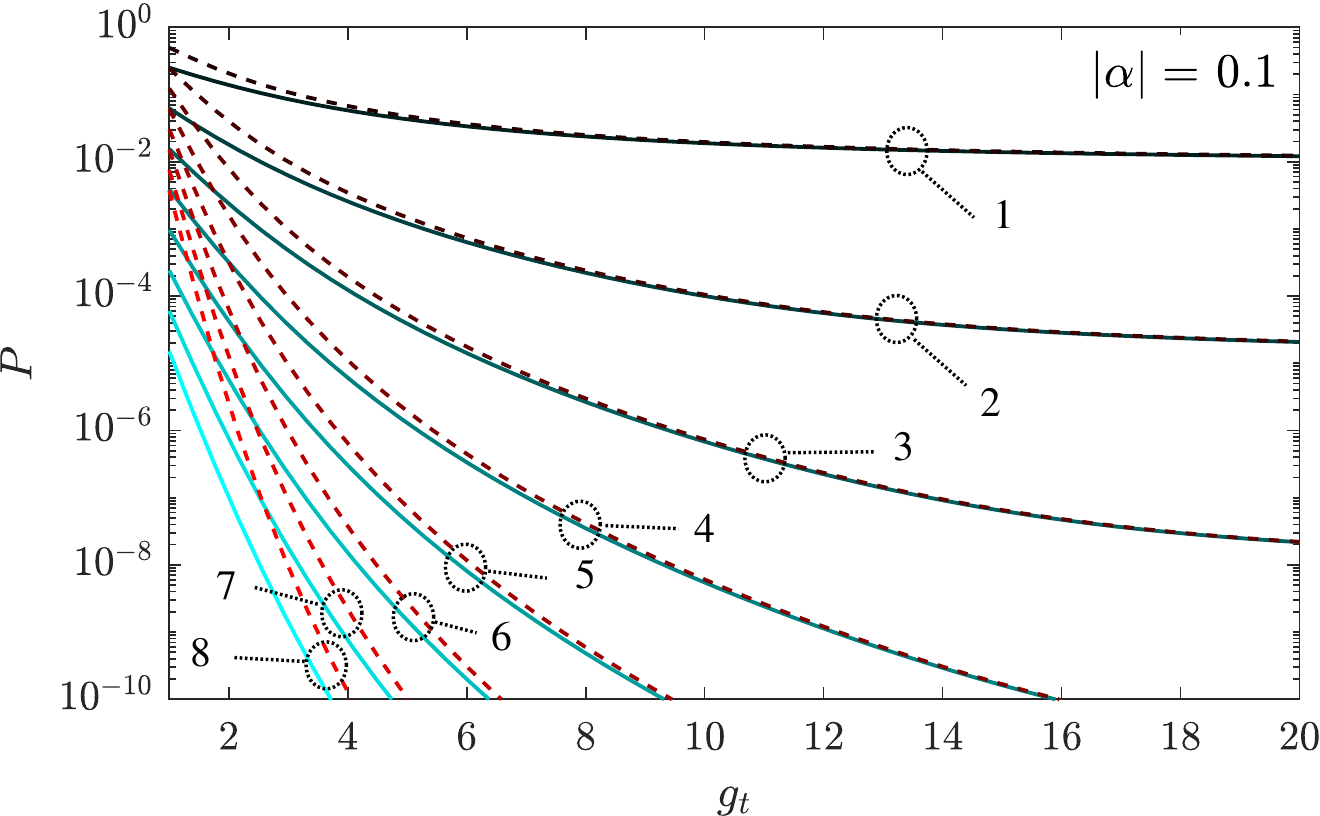}	
	\includegraphics[width=.32\linewidth]{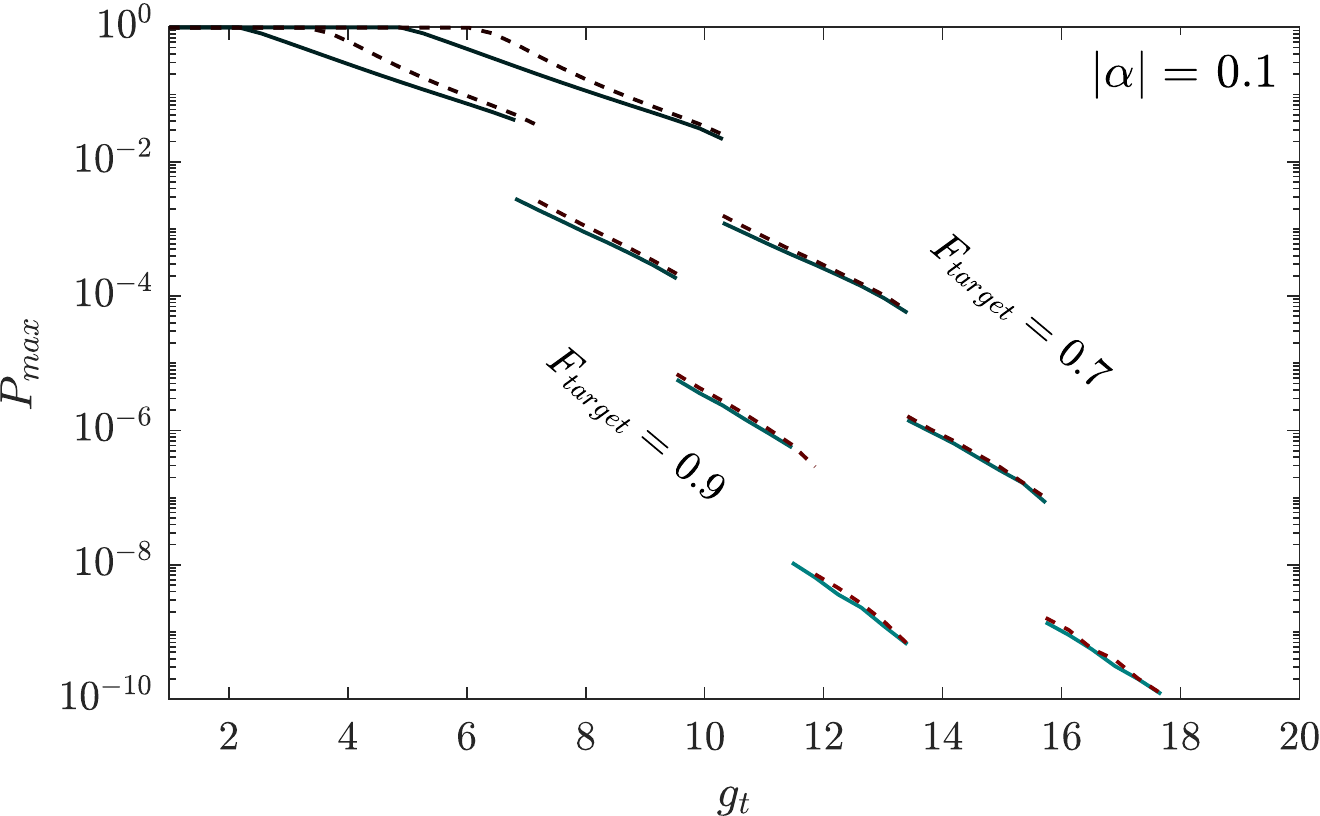}
	\\
	\includegraphics[width=.32\linewidth]{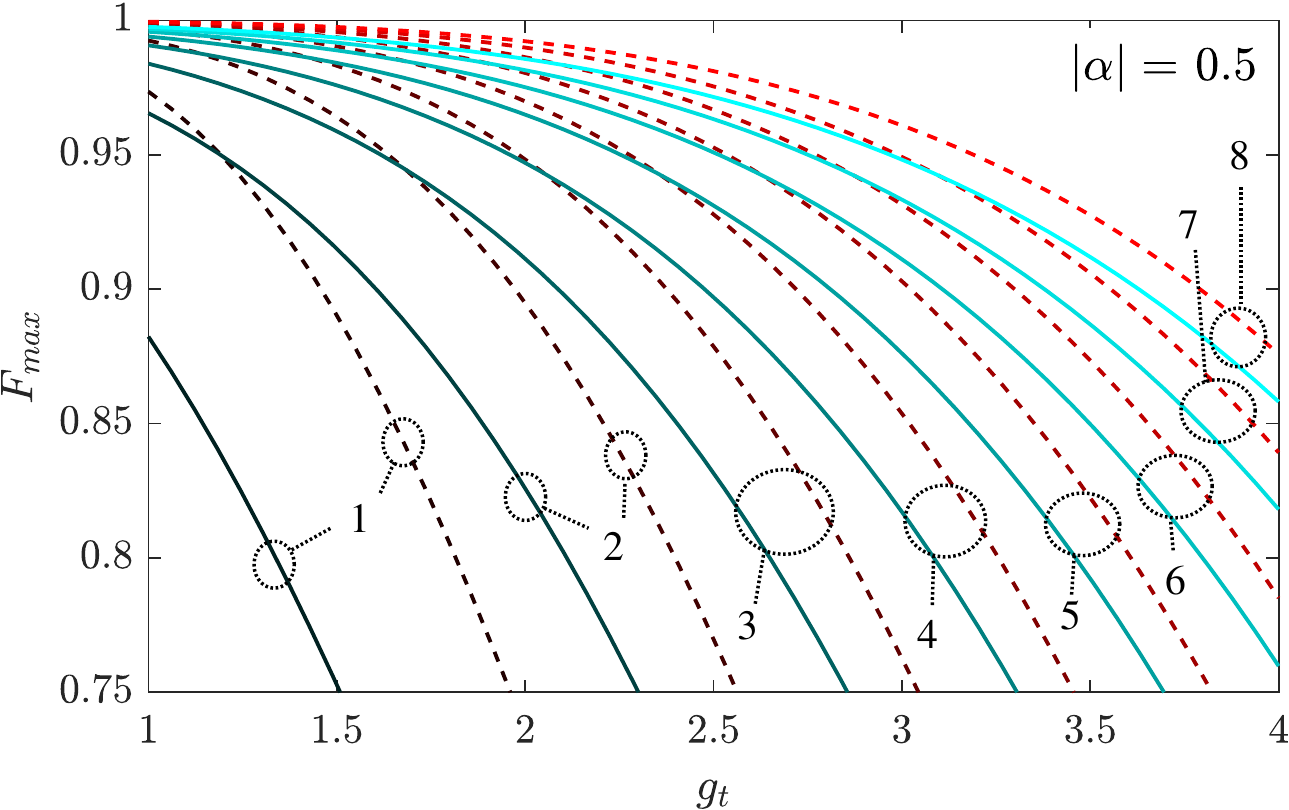}
	\includegraphics[width=.32\linewidth]{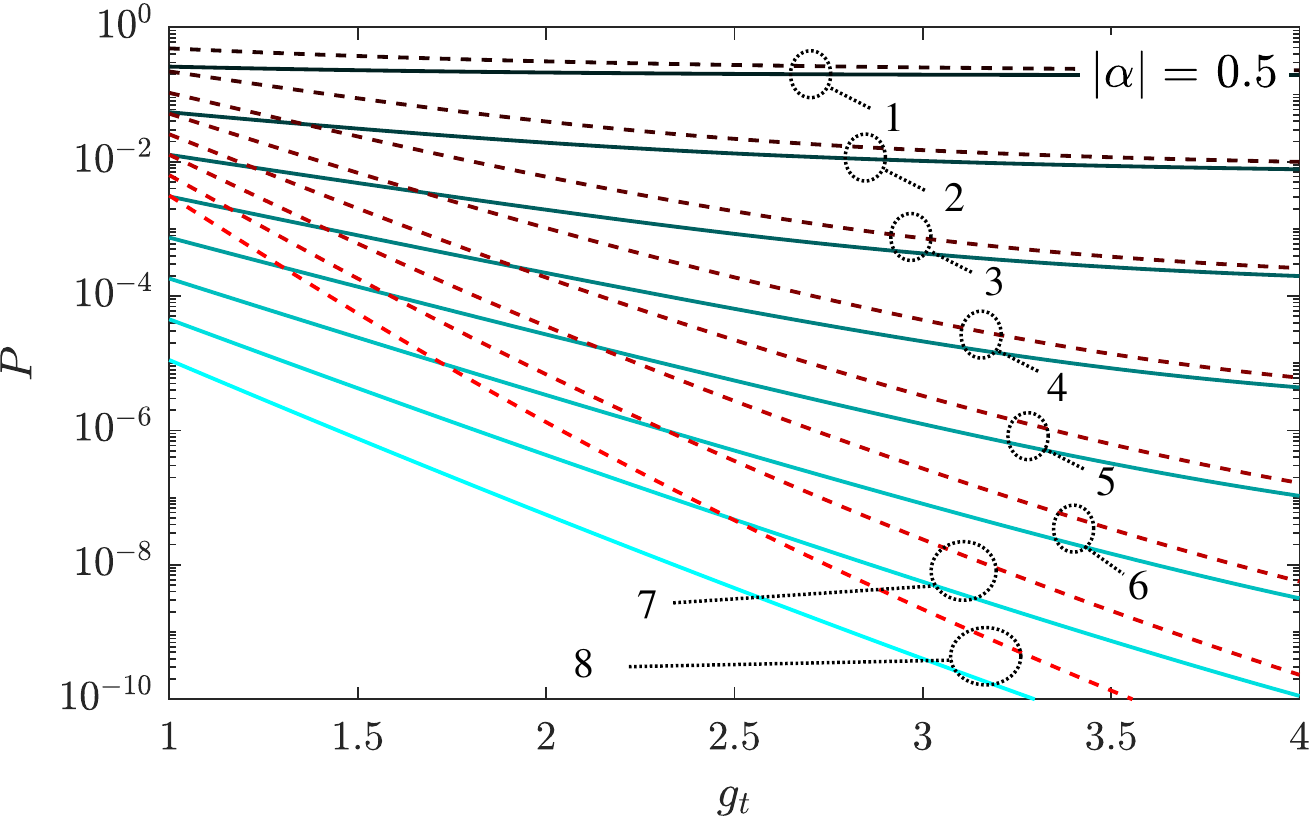}
	\includegraphics[width=.32\linewidth]{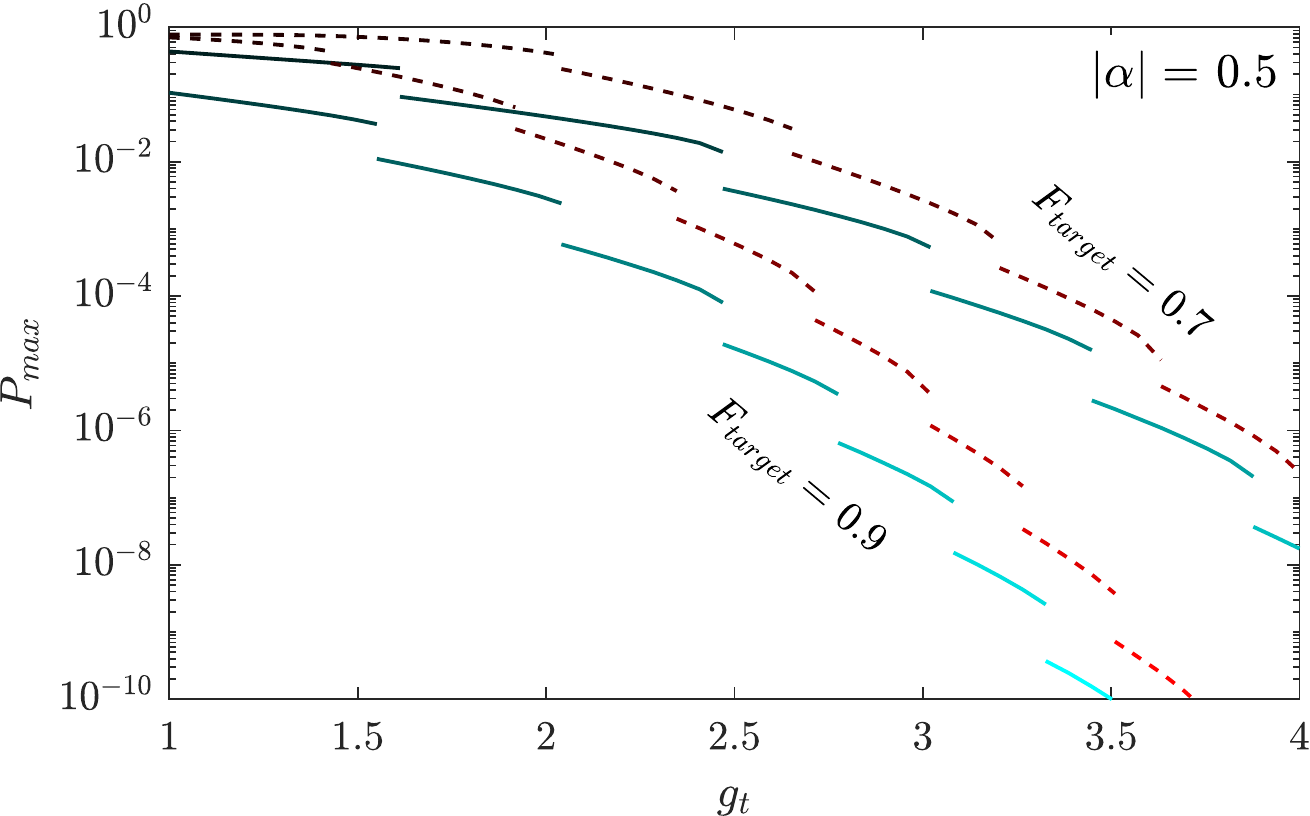}
	\caption{Left Column: The maximal achievable fidelity of the amplified states vs. the target amplification gain.
	The numbers connected to the curves indicate the cases for different $N$ (from $N=1$ to $N=8$).
	Middle Column: The corresponding success probabilities of the left column.
	Right Column: The maximal achievable success probabilities that guarantee certain levels of fidelity $F_{\text{obj}}$. 
	For different $F_{\text{obj}}$ the curves in the right column are disjointed since the success probabilities are maximized in terms of $N$.
	In all figures the dashed curves correspond to the QS-NLA and the solid curves correspond to the PC-NLA.}
	\label{fig:cohfidelity}
\end{figure*}

Since the success probabilities for both QS-NLA and PC-NLA vanish as $N$ grows, for practical purposes it is only meaningful to compare the QS-NLA and the PC-NLA when $N$ is finite.

In the Fock basis, the transformation of the QS-NLA can be expressed as \cite{dias2017quantum}
\begin{equation}\label{eq:mqs}
\hat{M}_s = \sum_{n=0}^N\sqrt{T}^N \frac{N!}{(N-n)!N^n}\sqrt{\frac{1-T}{T}}^n \ket{n}\bra{n}.
\end{equation}
We note $\hat{M}_s$ will truncate $\ket{n}, \forall n>N$.
In the Fock basis, the PC-NLA implements the transformation
\begin{equation}\label{eq:nlapcn}
\begin{aligned}
&\hat{M}_c=\sum_{n=0}^\infty
\sqrt{T}^{N+n}
\sum_{k=0}^{N}
\begin{pmatrix}
N\\k
\end{pmatrix}
\frac{n!}{(n-N+k)!}N^{k-N}p^{N-k}
\ket{n}\bra{n},
\end{aligned}
\end{equation}
where $p=\frac{T-1}{T}$.
The derivation for Eq.~(\ref{eq:nlapcn}) can be found in Appendix \ref{ap:2}.

We now numerically calculate (for finite $N$) the fidelity between a target coherent state, $\ket{g_t\alpha}$ and the state after amplification, $\ket{\psi}_{out}$. Here, $g_t$ is the target amplification gain. Note, for the PC-NLA the target coherent state is $\ket{-g_t\alpha}$, due to the global phase shift.
Since both $\ket{g_t\alpha}$ and $\ket{\psi}_{out}$ are pure states, the fidelity is simply the overlap of the two states
\begin{equation}
F=\left|\braket{g_t\alpha|{\psi}_{out}}\right|^2.
\end{equation}
The results are shown in Fig.~\ref{fig:cohfidelity}.
In the left column of Fig.~\ref{fig:cohfidelity} we vary the value of $g_t$, and for each $g_t$ we find the $T$ that maximizes $F$.
The maximization on $F$ for the two NLAs are carried out independently.
We can see for each NLA, at fixed $g_t$, the maximal fidelity increases as $N$ grows.
To achieve a certain level of fidelity, input coherent states with a higher amplitude require larger $N$.
The success probability for the NLAs when achieving the maximal fidelity are illustrated in the middle column of Fig.~\ref{fig:cohfidelity}.
When $N$ and $g_t$ are both fixed, the QS-NLA always provides higher maximal fidelity at larger success probabilities than the PC-NLA.
In the right column of Fig.~\ref{fig:cohfidelity} we investigate the maximal success probabilities that guarantee a certain level of fidelity,
\begin{equation}\label{eq:p_max_f}
P_{\text{max}}=\max_{T,\, N} \{P\}\, \text{s.t.}\, F\geq F_{\text{target}},
\end{equation}
where $F_{\text{target}}<1$ is the target fidelity.
From the results shown we can conclude at fixed $g_t$ the QS-NLA always has a higher success probability with regard to achieving a target fidelity.
The difference of the probabilities of the two NLAs is insignificant when $|\alpha|$ is small.
{All these conclusions also hold for $N=1$.}

\section{Entanglement Distillation}\label{sec:compare}
\begin{figure*}
	\centering	
	\includegraphics[width=.32\linewidth]{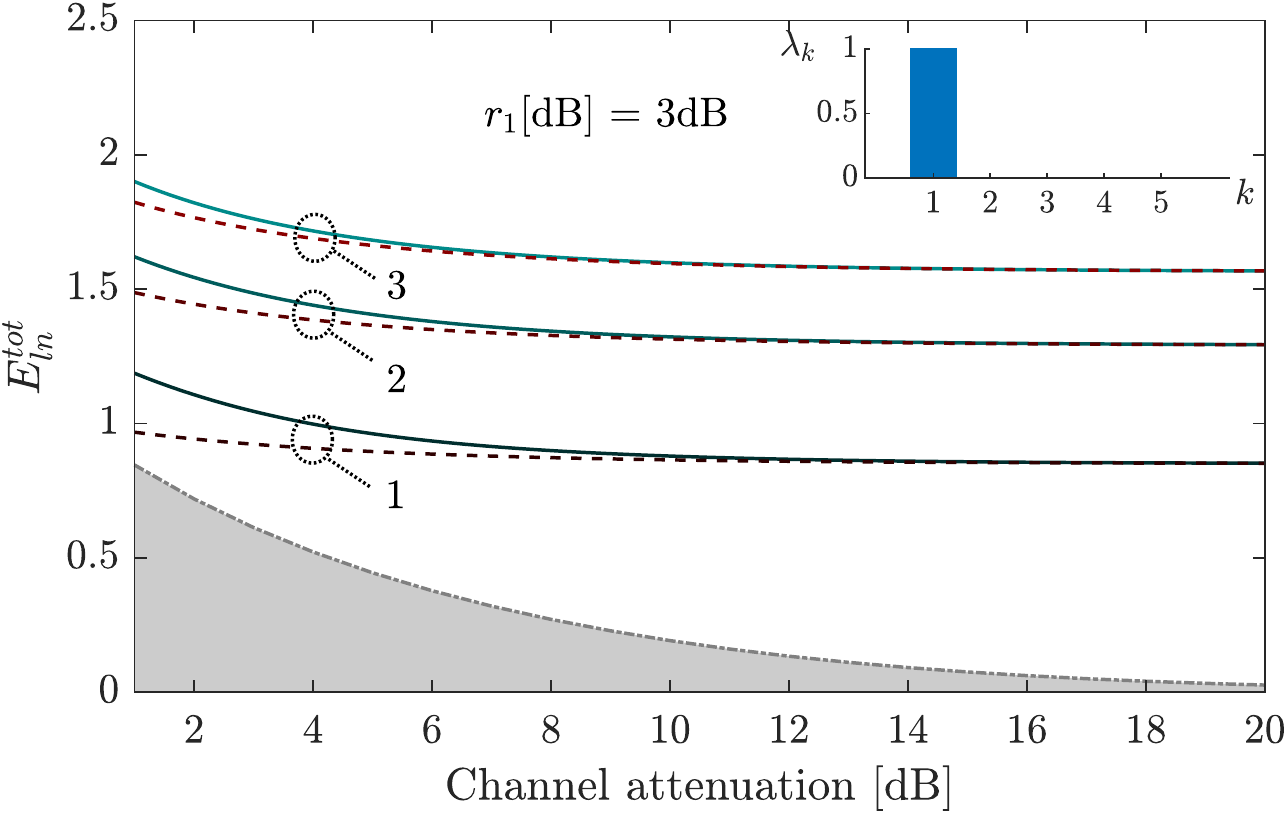}
	\includegraphics[width=.32\linewidth]{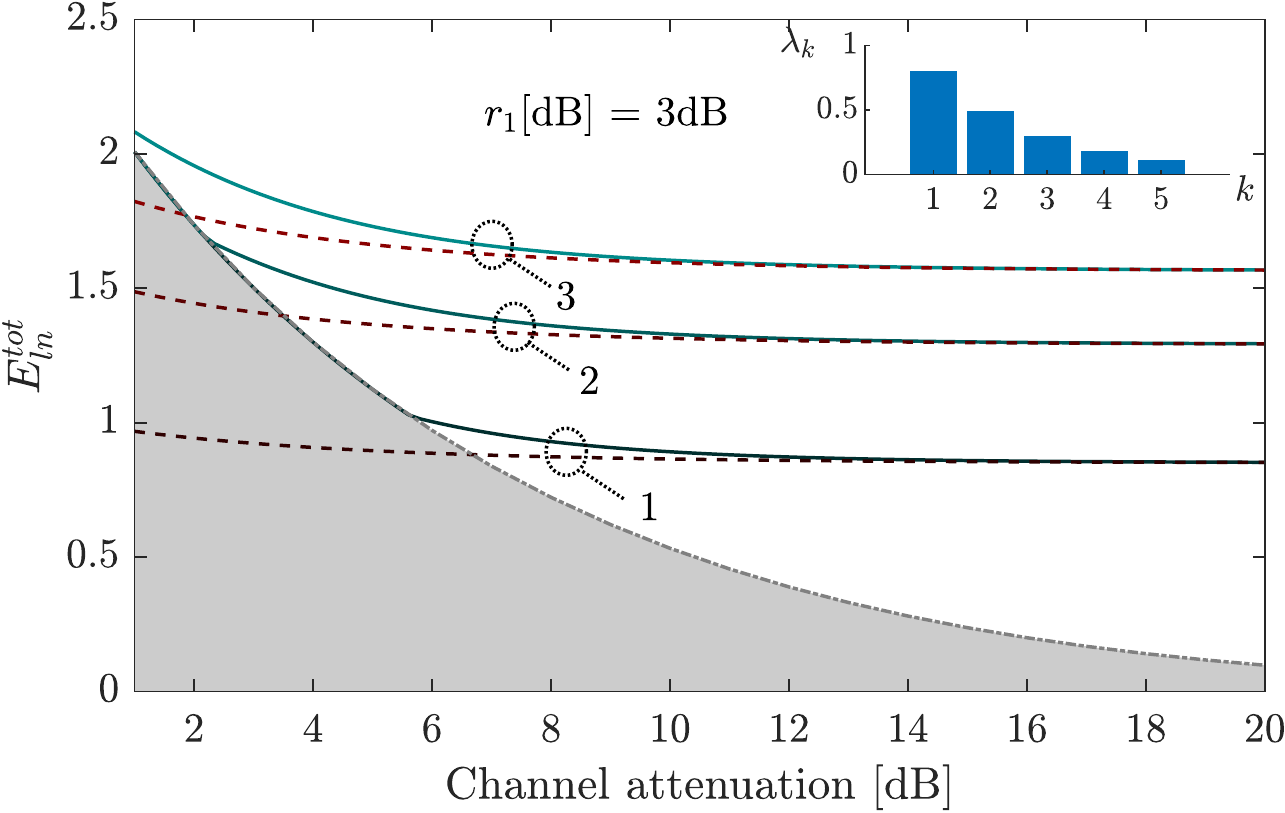}
	\includegraphics[width=.32\linewidth]{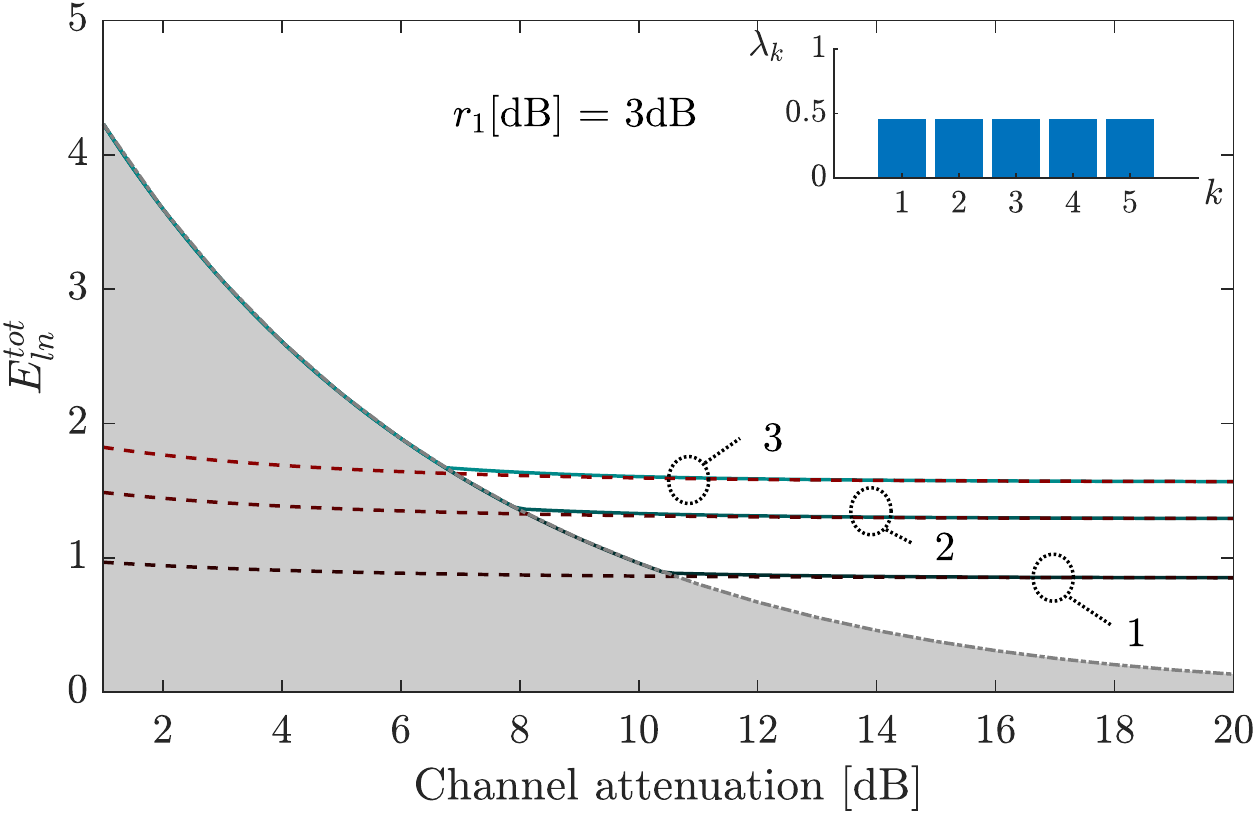}\\
	\includegraphics[width=.32\linewidth]{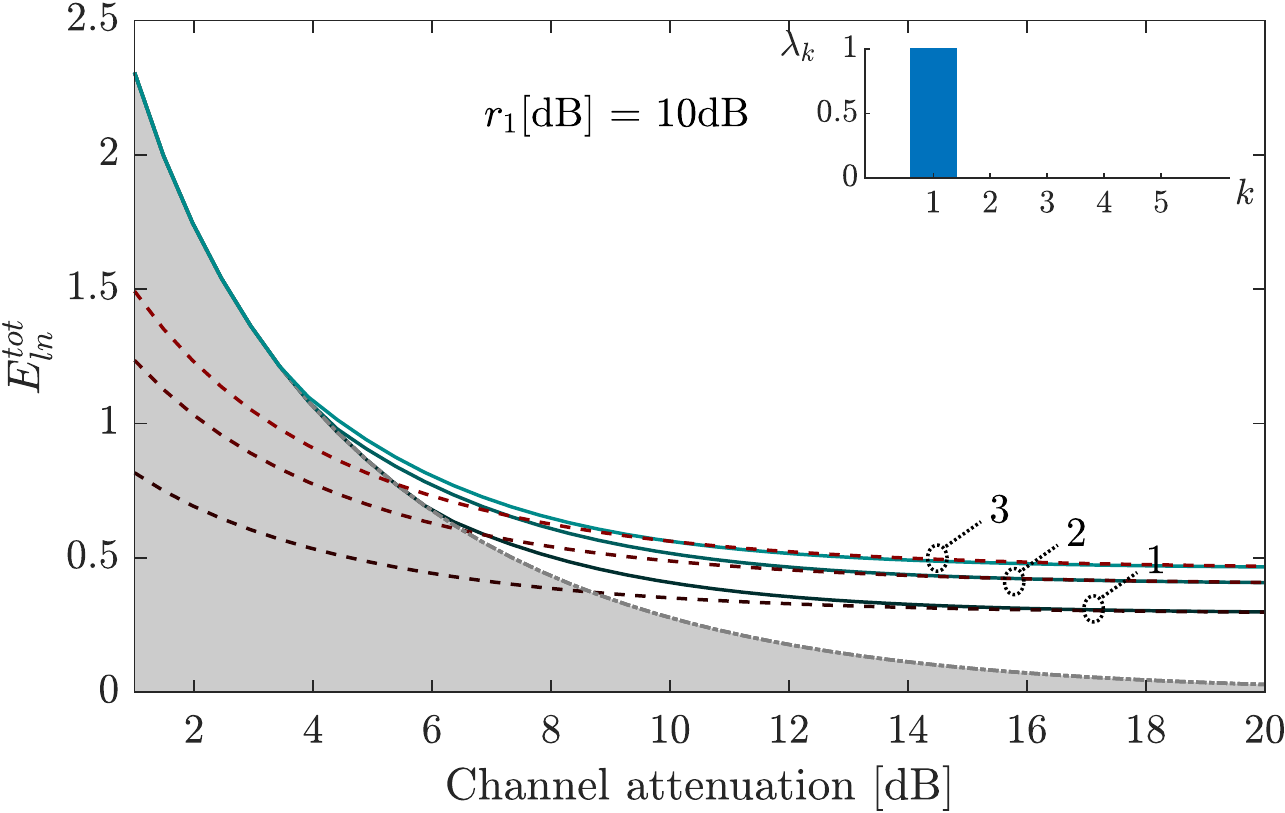}
	\includegraphics[width=.32\linewidth]{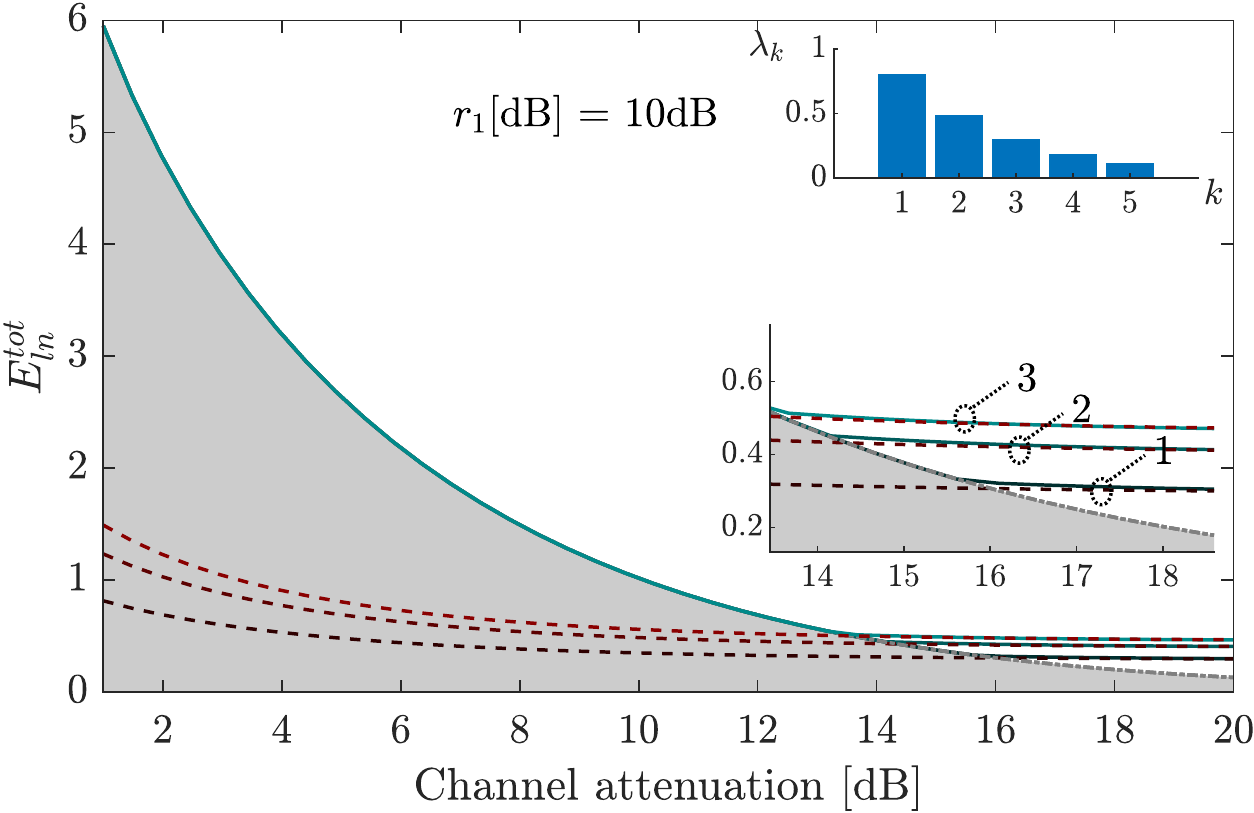}
	\includegraphics[width=.32\linewidth]{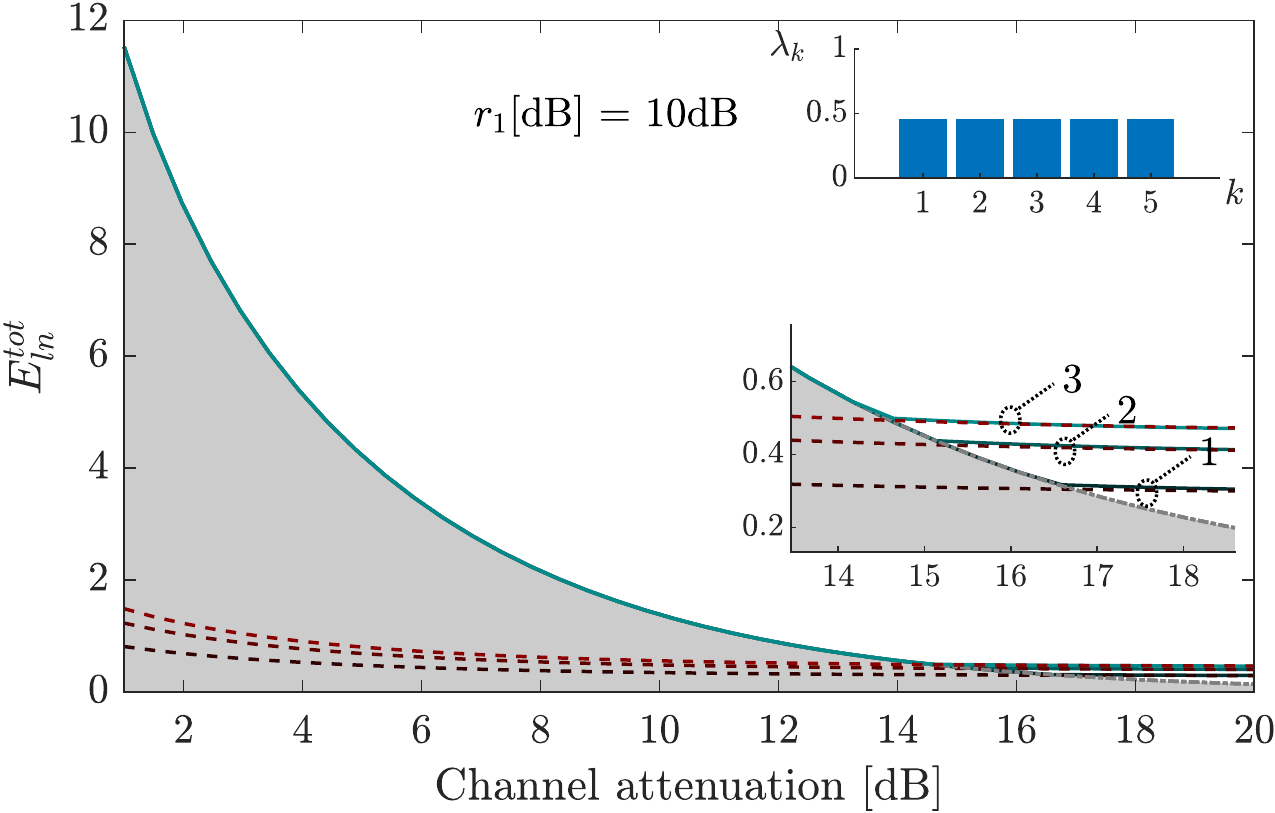}
	\caption{The maximized log-negativity of the amplified states vs. the channel attenuation.
		The dashed curves correspond to the QS-NLA and the solid curves correspond to the PC-NLA.
		The value of $N$ is indicated by the number connected to the curves.
		The edge of the gray area (also marked as dashed-dotted gray curve) represents the reference case without NLAs.
		The squeezing in dB is calculated by $r_1[\text{dB}]\approx 8.67r_1=8.67G\lambda_1$.
		The insets illustrate the supermode structure of the PDC state.}
	\label{fig:eln}
\end{figure*}

In this section we study the use of NLAs in the context of entanglement distillation over lossy channels.
We first briefly review the Parametric Down-Conversion (PDC) process, which is commonly used to generate entangled states.

In reality, the PDC process does not generate a single Einstein-Podolsky-Rosen (EPR) state with two entangled single-modes, but rather an ensemble of orthogonal EPR states each consisting of two entangled supermodes ({the orthogonalized Gram-Schmidt modes} \cite{de2014full}).
In the PDC process, a pump laser is first fed into a non-linear crystal. Two correlated beams, labeled $\bm{A}$ and $\bm{B}$, are then created.
Let $\hat{A}^\dagger_k$ and $\hat{B}^\dagger_k$ be the creation operators of the supermodes in beams $\bm{A}$ and $\bm{B}$, respectively, where we use the subscript $k\in\left\{1,2,...,\infty\right\}$ to index the supermodes (or equivalently, the EPR states).
The supermode operators satisfy the commutation relations $[\hat{A}_k,\hat{A}_{k'}^\dagger]=[\hat{B}_k,\hat{B}_{k'}^\dagger]=\delta_{kk'}$.
The output state of the PDC process, which we refer to as a PDC state, can be written as \cite{christ2012exponentially}
\begin{equation}\label{eq:PDCinSch}
\begin{aligned}
\ket{\text{PDC}}_{AB} &= \bigotimes_{k=1}^\infty \exp\left[G \lambda_{k}\left(\hat{A}^\dagger_k \hat{B}^\dagger_k  - \hat{A}_k \hat{B}_k\right) \right]\ket{0}\\
&=\bigotimes_{k=1}^\infty \ket{\text{EPR}_k},
\end{aligned}
\end{equation}
where
$G$ is the overall gain of the PDC process, the $\lambda_{k}$'s are normalized real-valued coefficients, and $r_k=G \lambda_{k}$ is the squeezing parameter for the $k$-th EPR state.

Consider a scenario where one beam of a PDC state is distributed over lossy channels while another beam is kept at the transmitter.
We assume the channel is frequency independent such that the supermode structure of the distributed beam is retained after passing through the channel.
We assume the number of orthogonal EPR states of the PDC state is $K=5$.\footnote{{In experiments $K$ can be considered as the number of frequency resolved bins for the specific detector used.}}

For the supermode structure of the PDC state we consider three scenarios.
For the first scenario, {the PDC state only contains one non-trivial EPR state}, i.e., $r_k\approx0,\, \forall k \neq 1$.
For the second scenario, the PDC state contains 5 EPR states with squeezing parameters ($r_1$ to $r_5$) following an exponentially decaying distribution.
For the third scenario, the PDC state contains 5 EPR states with the same amount of squeezing.

We assume an NLA is applied to the first supermode ($k=1$) of the received beam.
We first consider an amplification strategy where the supermode to be amplified is not filtered out (spatially separated from the other supermodes) before the amplification.
This strategy is commonly adopted for various multimode non-Gaussian operations, such as photon subtraction and photon addition \cite{averchenko2014nonlinear,averchenko2016multimode,walschaers2017statistical,walschaers2019mode,ra2020non}.
{We focus on this strategy also because in experiments the supermodes cannot be easily separated} \cite{huo2020direct}.

We use the QS-NLA as an example to explain how the NLA.
The received beam is fed into the QS-NLA and divided into $N$ beams by the $N$-splitter.
Parallel QS operations are then applied to all $N$ beams.
In each QS operation, the multimode single-photon detector, which implements a joint detection on the frequency bins,\footnote{The measurement of a supermode can be realized by a multi-pixel detection method (e.g. \cite{armstrong2012programmable,de2014full,plick2018violating,cai2020quantum}), in which the supermode is fanned out by a diffraction grating onto homodyne detectors, each having a resolution of one frequency bin (i.e., one pixel).
The joint single-photon detection on the frequency bins can be implemented by a similar method, in which the homodyne detectors are replaced with mode-non-selective single-photon detectors.} clicks if and only if a photon with the same multimode structure as the first supermode of the beam is detected.
This also means no photon in the rest of the supermodes is detected.
In this case, the transformation by the QS on the rest of the supermodes can be represented by
$\bra{0}\bra{0}
\textbf U(1/2)
\textbf U(T)
\ket{0}\ket{0}
=\ket{0}\bra{0}$.
The transformation by the QS-NLA on these supermodes is also $\ket{0}\bra{0}$ (i.e., a truncation to the vacuum state). 
The transformation by the QS-NLA on the first supermode is $\hat{M}_s $ as given by Eq.~(\ref{eq:mqs}).

Similarly, it can be shown when a PC-NLA ($\hat{M}_c $ as given by Eq.~(\ref{eq:nlapcn})) is applied to the first supermode the rest of the supermodes will be attenuated. The attenuation can be described by an operator $\sqrt{T}^{\hat{A}^\dagger \hat{A}}$.

We compare the QS-NLA and the PC-NLA in terms of the maximal achievable  log-negativity of the ensemble of the state after the amplification, which is defined by
\begin{equation}
E_{\text{ln}}^{\text{tot}}=\max_{T\in (0,1)}\{\sum_{k=1}^{5} E_{\text{ln}}[\rho_k]\},
\end{equation}
where $\rho_k$ is the density operator for the $k$-th distributed (and possibly amplified) EPR state.
The log-negativity for each $\rho_k$ is defined as
\begin{equation}
E_{\text{ln}}[\rho_k]=\log_2[1+2\epsilon(\rho_k)],
\end{equation}
where $\epsilon(\rho)$ stands for the absolute value of the sum of negative eigenvalues of the partially transposed $\rho$.
The maximization of $E_{\text{ln}}^{\text{tot}}$ is performed on the transmissivity $T$ of the beam-splitters in the NLAs.

The results are illustrated in Fig.~\ref{fig:eln}. 
For the first scenario (only one {non-trivial} EPR state), at fixed $N$, the PC-NLA provides larger $E_{\text{ln}}^{\text{tot}}$ than the QS-NLA when the channel attenuation is below certain thresholds (less loss).
The two NLAs show similar performances when the attenuation is above these thresholds.
Interestingly, both NLAs can achieve a certain level of $E_{\text{ln}}^{\text{tot}}$,  independent of the channel attenuation level.
The truncation effect of the QS-NLA degrades $E_{\text{ln}}^{\text{tot}}$ significantly when the initial squeezing of the EPR state is large.
For the second and third scenarios (more than one non-trivial EPR state), the increase of $E_{\text{ln}}$ from the amplified EPR state is negated by the decrease of $E_{\text{ln}}$ from the other EPR states, making the thresholds above which the NLAs can enhance $E_{\text{ln}}^{\text{tot}}$ higher than the first scenario.

For the amplification strategy where the supermode to be amplified is first filtered out before the amplification, the rest of the supermodes remain unchanged after the amplification process.
Under this strategy the performances of each NLA will be less dependent on the supermode structure of the PDC states.

In summary, the NLAs can only enhance the entanglement when the channel attenuation is above some thresholds (i.e. larger loss).
These thresholds depend on the supermode structure, the squeezing levels, and the number of amplifying units (QS or PC) $N$ in the NLAs.
In general, the thresholds increase as the entanglement of the initial PDC states increases.

\section{Comparison of the PC-NLA and the cascaded processing of PC}\label{sec:parcas}
\begin{figure}
	\centering
	\includegraphics[width=0.8\linewidth]{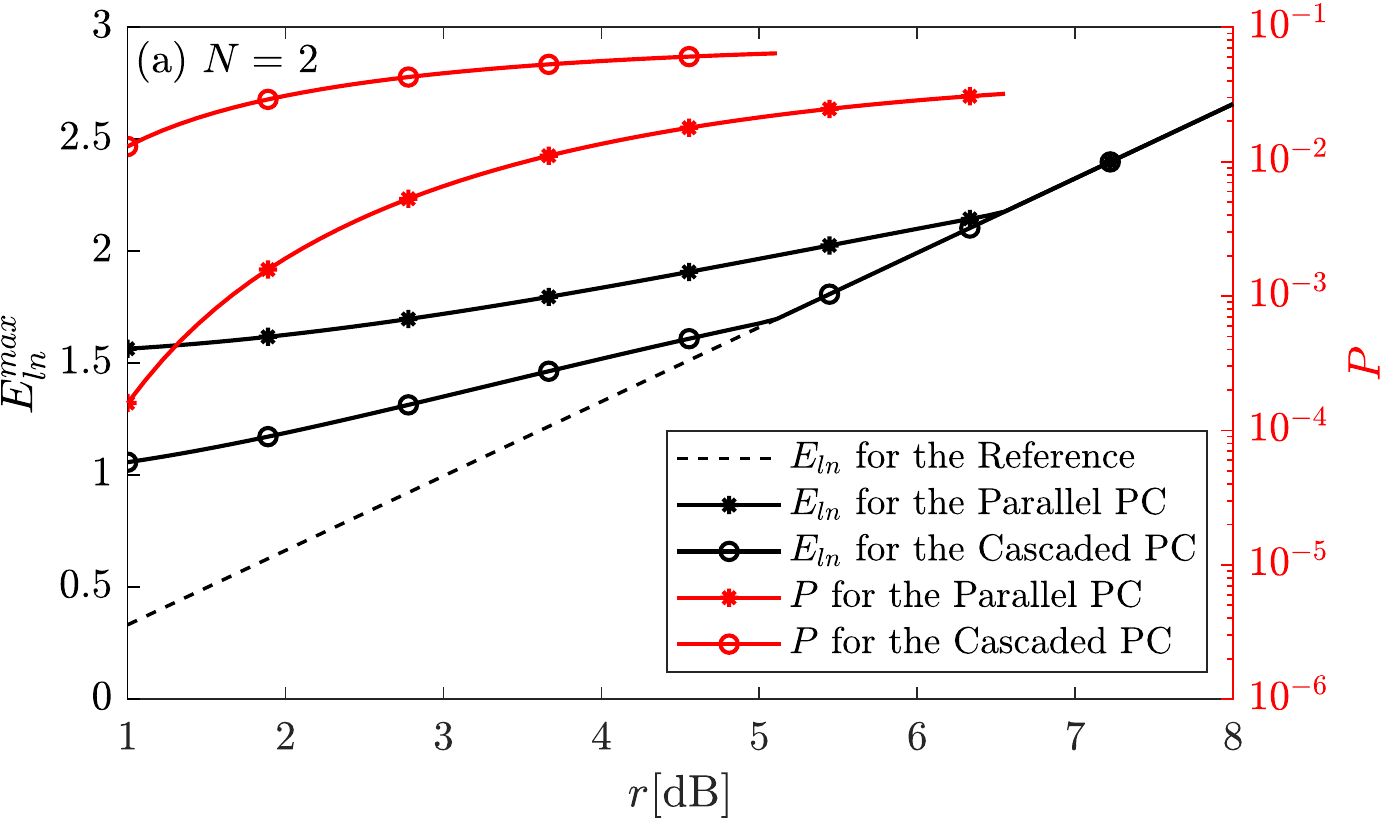}\\
	\includegraphics[width=0.8\linewidth]{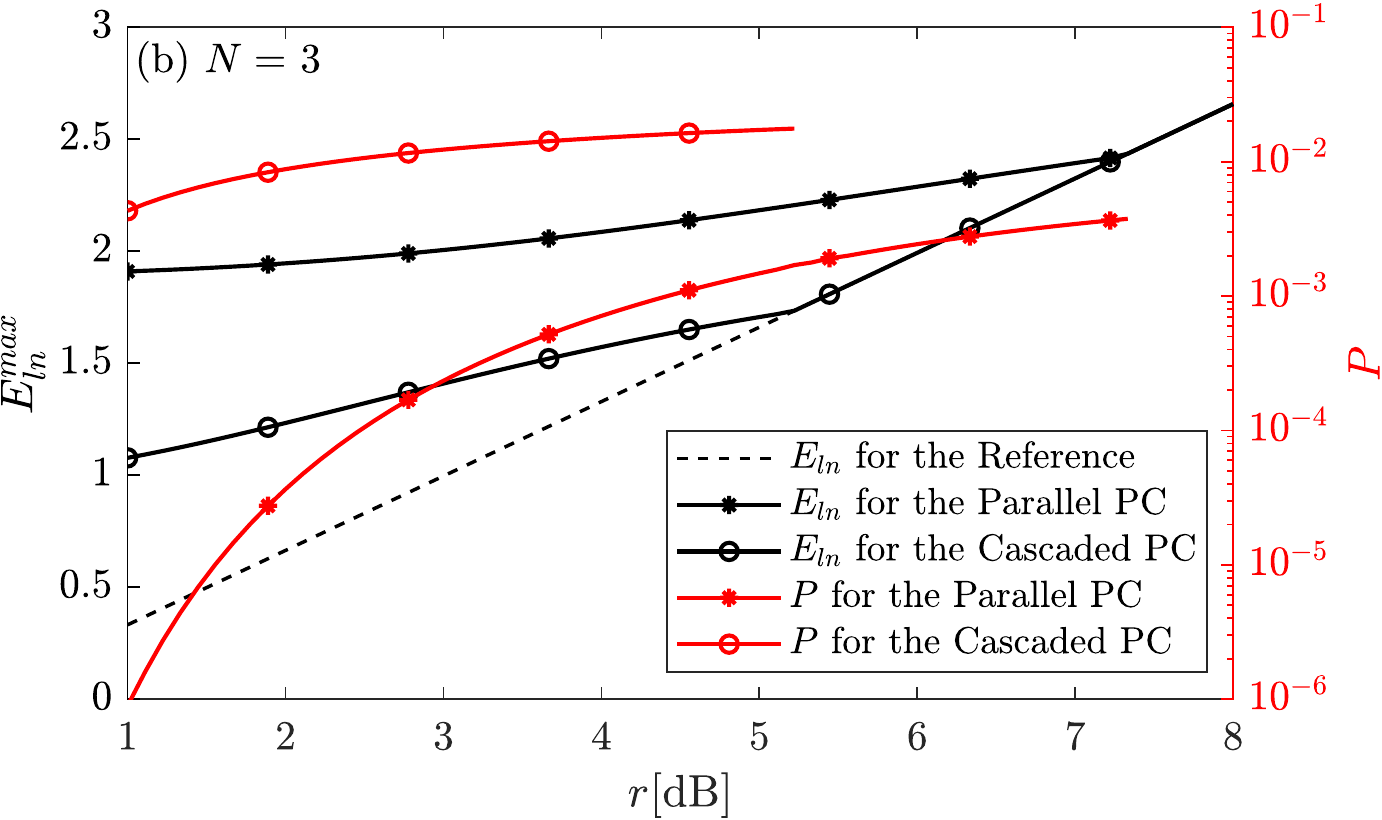}
	\caption{The maximal log-negativity $E_{\text{ln}}^{\text{max}}$ (black curves) and the success probability $P$ (red curves) against $r[\text{dB}]$ for (a) $N=2$ and (b) $N=3$.
		The black dashed curve represents the reference case without any PC operations.}
	\label{fig:ppccpcn}
\end{figure}

\begin{figure}
	\centering
	\includegraphics[width=0.8\linewidth]{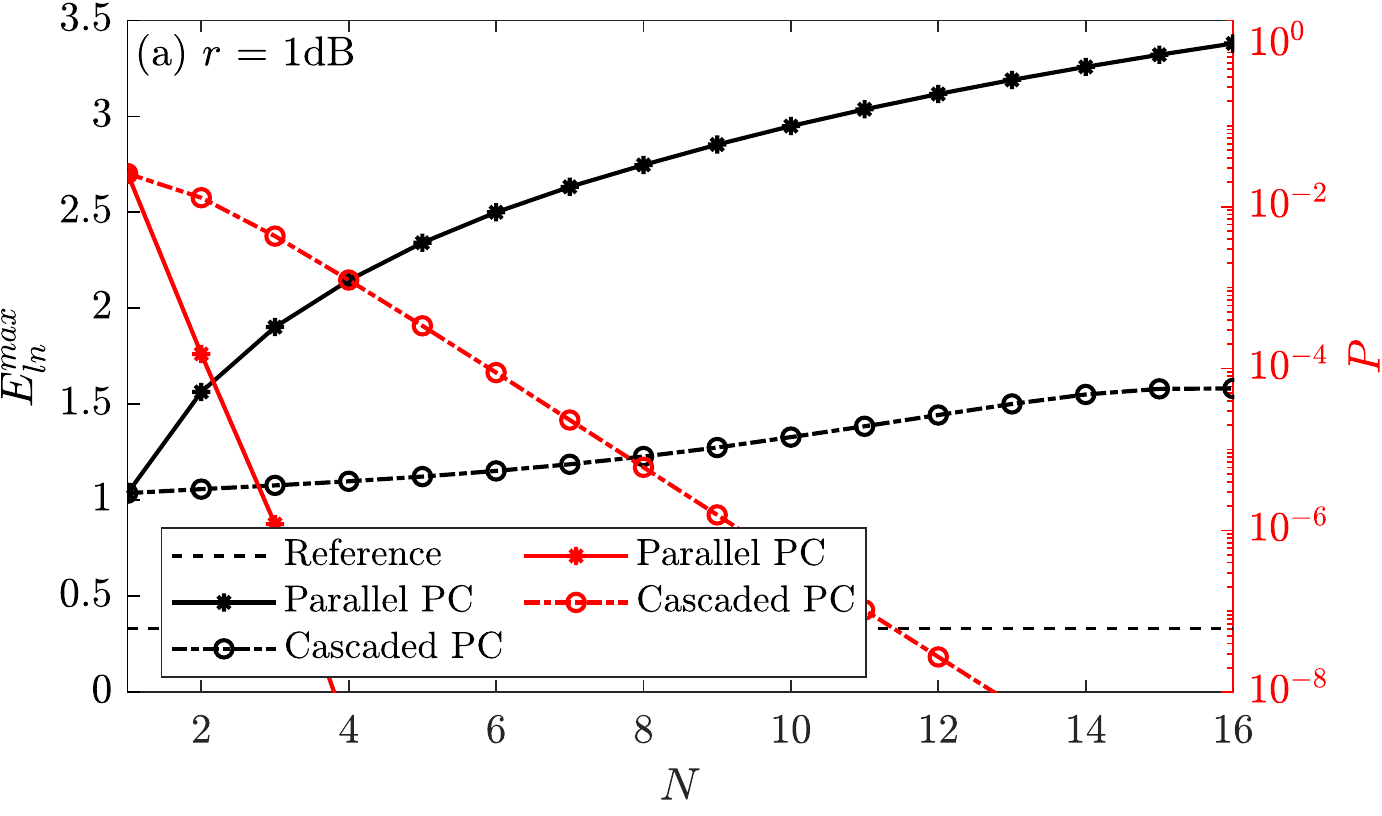}\\
	\includegraphics[width=0.8\linewidth]{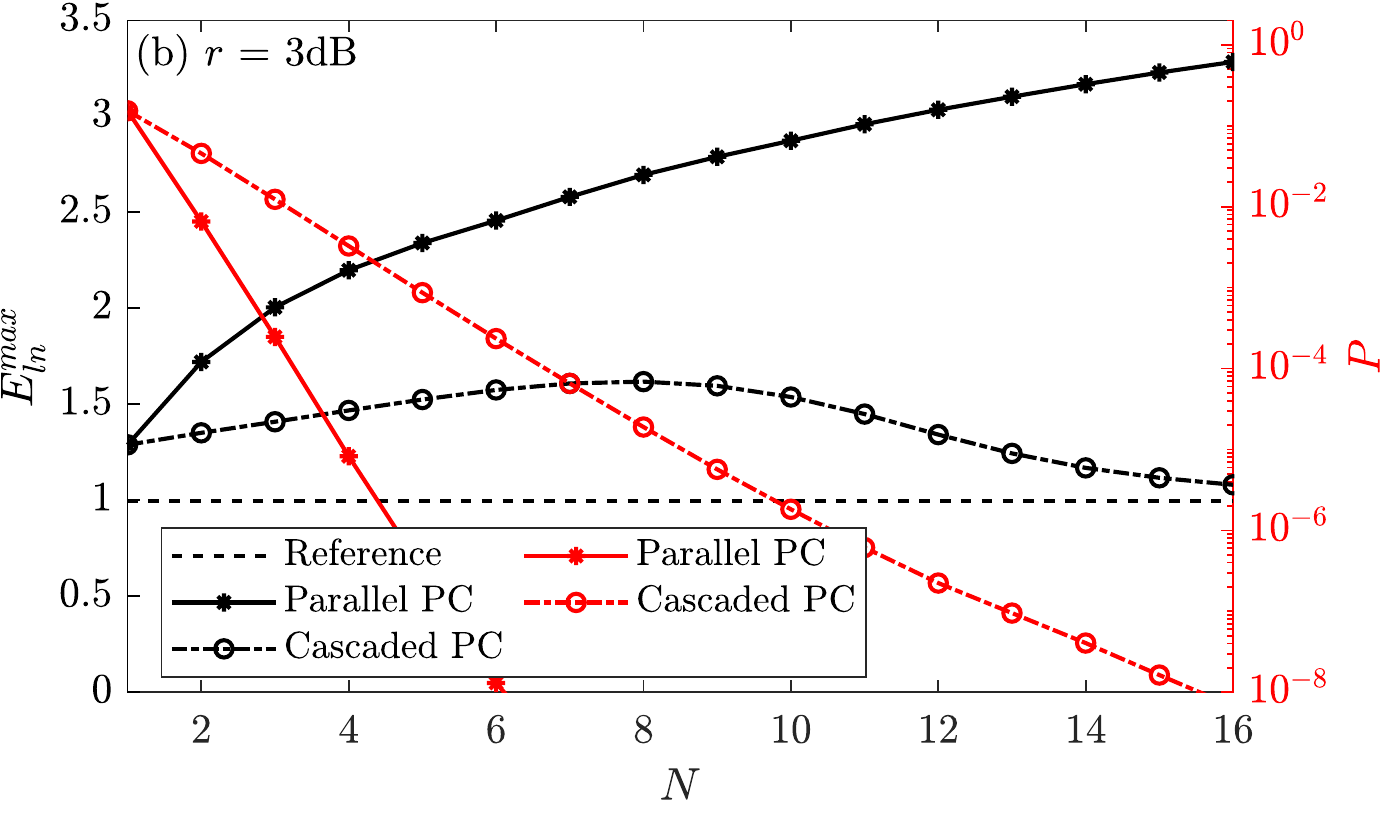}
	\caption{The maximal log-negativity $E_{\text{ln}}^{\text{max}}$ (black curves) and the success probability $P$ (red curves) against $N$ for (a) $r=1$dB and (b) $r=3$dB.
		The black dashed curve represents the reference case without any PC operations.}
	\label{fig:ppccpcnr3}
\end{figure}

In this section, we compare the performance of our PC-NLA (the parallel processing of PC) and the cascaded processing of PC proposed in \cite{mardani2020continuous} in the context of entanglement distillation of EPR states.
To better study the impact of the two processes on EPR states, we restrict ourselves to the scenario where the PDC state only contains one EPR state.
{We note our conclusions in this section will also apply to the scenario where the PDC state contains more than one EPR state.}

We first generalize the cascaded PC to the multimode setting so as to compare the two processes.
Recall the PC operation, which we repeat here for completeness, can be represented by
\begin{equation}\label{eq:pcmulti}
\hat{R}=\sqrt{T}\left(-\frac{1-T}{T}\hat{A}^\dagger\hat{A}+1\right)\sqrt{T}^{\hat{A}^\dagger\hat{A}}.
\end{equation}
The cascaded PC, which contains $N$ repetitions of the same PC operation, can then be represented by
\begin{equation}\label{eq:pccasN}
\hat{R}^N=\sqrt{T}^N\left(-\frac{1-T}{T}\hat{A}^\dagger\hat{A}+1\right)^N\sqrt{T}^{N\hat{A}^{\dagger}\hat{A}},
\end{equation}
where we have used the fact that $\sqrt{T}^{\hat{A}^{\dagger}\hat{A}}$ and $\hat{A}^{\dagger}\hat{A}$ commutes.

In Fig.~{\ref{fig:ppccpcn}} and Fig.~{\ref{fig:ppccpcnr3}} we compare the log-negativity for the two processes in the absence of channel losses.
For the two processes the beam-splitter transmissivity $T$ is optimized individually so as to maximize $E_{\text{ln}}$.
In Fig.~{\ref{fig:ppccpcn}} the maximal log-negativity, $E_{\text{ln}}^{\text{max}}$, is plotted against the initial squeezing $r$ of the EPR state.
In Fig.~{\ref{fig:ppccpcnr3}} we plot $E_{\text{ln}}^{\text{max}}$ against the number of PC operations in each process.
From the figures we can see the parallel PC can achieve much higher $E_{\text{ln}}^{\text{max}}$ than the cascaded PC.
The difference between $E_{\text{ln}}^{\text{max}}$ for the two processes becomes more significant as $N$ grows.
However, there is a trade-off between the success probabilities and the achievable $E_{\text{ln}}^{\text{max}}$.
The cascaded PC can be orders of magnitude more successful than the parallel PC in terms of probability at the price of a drop in $E_{\text{ln}}^{\text{max}}$.

We attempt to better explain why for the cascaded PC the entanglement increases insignificantly (and even decreases) as $N$ grows.
From Eq.~(\ref{eq:pcmulti}) it can be observed the operator for PC can be factorized into two terms, namely, $\sqrt{T}\left(-\frac{1-T}{T}\hat{A}^\dagger\hat{A}+1\right)$ and $\sqrt{T}^{\hat{A}^{\dagger}\hat{A}}$.
The latter term is the operator for a noiseless linear attenuator, which only decreases the entanglement of an EPR state \cite{yang2012continuous}.
When multiple cascaded PC operations are applied, the attenuation term scales as $\sqrt{T}^{N\hat{A}^{\dagger}\hat{A}}$ (the rightmost term in Eq.~(\ref{eq:pccasN})). The detrimental impact of the attenuation term will eventually negate the entanglement.
For the parallel PC the attenuation term is independent of $N$.
Therefore, the entanglement grows steadily as $N$ grows.

\section{Conclusions}\label{sec:conclusion}
In this work, we proposed for the first time an NLA that uses parallel processing of PC, the PC-NLA.
We also constructed a multimode version of an existing NLA that uses QS, namely, the QS-NLA.
We showed that when applied to a range of coherent states, the PC-NLA, which can be built with much simpler linear optics, is compatible with the QS-NLA.
In the context of entanglement distillation of PDC states, we found that both NLAs can enhance the entanglement when the channel attenuation is above certain thresholds.
Distinct from the single-mode NLA analysis, these thresholds largely depend on the supermode structure of the PDC states.
An interesting finding is that the two NLAs can maintain certain levels of entanglement, independent of the channel attenuation level.
We also compare the PC-NLA with the cascaded processing of PC, showing that the PC-NLA can distill more entanglement, albeit at lower success probabilities.
Our results will be important for next-generation real-world implementations of multipartite quantum information applications that utilize broadband pulses of lights.

\begin{acknowledgments}
Mingjian He is partially supported by the China Scholarship Council.
\end{acknowledgments}

\appendix
\section{The derivation for Eq.~(\ref{eq:multiQS})}\label{ap:1}
In this section, for conciseness we omit the subscripts that label the spatial modes ($a$, $b$, $c$, etc.) but keep the subscripts that index the single-modes ($m\in\left\{1,2,...,\infty\right\}$).
From Eq.~(\ref{eq:multiQS}) it follows that
\begin{equation}\label{eq:multiQSdetail1}
\begin{aligned}
\hat M=&\bra{0}_C\bra{1}_A
\textbf U_{AC}(T_1)
\textbf U_{BC}(T_2)
\ket{1}_B\ket{0}_C\\
=&
\sum_{m',m''}\gamma_{m'}^* \gamma_{m''}
\bigotimes_{m}
\bra{0}_{{m}}
\bra{\delta_{m,m'}}_{{m}}
U_m(T_1)
U_m(T_2)
\ket{\delta_{m,m''}}_{{m}}
\ket{0}_{{m}}
\\
=&\sum_{m'}
\gamma_{m'}^* \gamma_{m'}
\hat{M}_{m'}^{(11)}
\bigotimes_{m, m\neq m'}
\hat{M}_{m}^{(00)}\\
&+
\sum_{m',m'',m'\neq m''}
\gamma_{m'}^* \gamma_{m''}
\Big[\hat{M}_{m'}^{(10)}
\otimes
\hat{M}_{m''}^{(01)}
\Big]
\bigotimes_{m,m\neq m',m\neq m''}
\hat{M}_{m}^{(00)},
\end{aligned}
\end{equation}
where $(\cdot)^*$ stands for the complex conjugate, and
\begin{equation}
\hat{M}_{m}^{(n_1n_2)}:=
\bra{0}_{{m}}
\bra{n_1}_{{m}}
U_m(T_1)
U_m(T_2)
\ket{n_2}_{{m}}
\ket{0}_{{m}},
\end{equation}
for $n_1,n_2\in{\left\lbrace0,1\right\rbrace}$.
After some algebraic manipulations, we find
\begin{equation}\label{eq:multiQSdetail2}
\begin{aligned}
\hat{M}_{m}^{(11)}=&\sqrt{T_1T_2}\ket{0}_m\bra{0}_m+\sqrt{(1-T_1)(1-T_2)}\ket{1}_m\bra{1}_m\\
\hat{M}_{m}^{(00)}=&\ket{0}_m\bra{0}_m\\
\hat{M}_{m}^{(10)}=&\sqrt{1-T_1}\ket{0}_m\bra{1}_m\\
\hat{M}_{m}^{(01)}=&\sqrt{1-T_2}\ket{1}_m\bra{0}_m
\end{aligned}
\end{equation}
Substituting Eq.~(\ref{eq:multiQSdetail2}) into Eq.~(\ref{eq:multiQSdetail1}) it follows that
\begin{equation}
\begin{aligned}
\hat{M}=&\sqrt{T_1T_2}
\bigotimes_{m}\ket{0}_m\bra{0}_m\\
&+
\sum_{m'} \gamma_{m'}^* \gamma_{m'} \sqrt{(1-T_1)(1-T_2)} \\
&\quad\bigotimes_{m} \ket{\delta_{m,m'}}_m \bra{\delta_{m,m'}}_m\\
&+
\sum_{m',m'',m'\neq m''}
\gamma_{m'}^* \gamma_{m''} \sqrt{(1-T_1)(1-T_2)} \\
&\quad\bigotimes_{m} \ket{\delta_{m,m''}}_m \bra{\delta_{m,m'}}_m\\
=&\sqrt{T_1T_2}\ket{0}_B\bra{0}_A+\sqrt{(1-T_1)(1-T_2)}\ket{1}_B\bra{1}_A.
\end{aligned}
\end{equation}
Similarly, we can also show that
\begin{equation}
\begin{aligned}
\hat M'=&\bra{1}_C\bra{0}_A
\textbf U_{AC}(T_1)
\textbf U_{BC}(T_2)
\ket{1}_B\ket{0}_C\\
=&\sqrt{(1-T_1)T_2}\ket{0}_B\bra{0}_A-\sqrt{(1-T_2)T_1}\ket{1}_B\bra{1}_A.
\end{aligned}
\end{equation}
When $T_1=1/2$, $\hat{M}$ and $\hat M'$ only differ in a phase shift.

\section{The derivation for Eq.~(\ref{eq:nlapcn})}\label{ap:2}
In this section, for conciseness we omit the subscripts for the supermodes.
We consider an $N$-splitter that implements the following transformation
\begin{equation}
[\hat{A}_1,\hat{A}_2,...,\hat{A}_N]^T_\text{out}=U[\hat{A}_1,\hat{A}_2,...,\hat{A}_N]^T_\text{in},
\end{equation}
where $U$ is an $N$-by-$N$ unitary matrix whose entries satisfy $u_{1j}=u_{i1}=1/\sqrt{N},\,\forall i,j \in \{1,2,...,N\}$, and $\hat{A}_1$ to $\hat{A}_N$ are the annihilation operators of the supermodes in the $N$ paths.
For an input Fock state $\ket{n}$, the ensemble state after the first $N$-splitter can be written as
\begin{equation}
\begin{aligned}
\ket{\psi'}&=\frac{N^{-\frac{n}{2}}\left(\hat{A}_1^\dagger+\hat{A}_2^\dagger+\cdots+\hat{A}_N^\dagger\right)^n}
{\sqrt{n!}}\ket{0}\\
&=
\frac{N^{-\frac{n}{2}}}
{\sqrt{n!}}
\sum_{n_i\ge 0,n_1+n_2+\cdots+n_N=n}\\
&\quad\begin{pmatrix}
n\\n_1,n_2,...,n_N
\end{pmatrix}
\quad\prod_{i=1}^NA_i^{\dagger n_i}
\ket{0},
\end{aligned}
\end{equation}
where
\begin{equation}
\begin{pmatrix}
n\\n_1,n_2,...,n_N
\end{pmatrix}
=\frac{n!}{n_1!n_2!\cdots n_N!}.
\end{equation}
The state after the PC operations can be written as
\begin{equation}
\begin{aligned}
\ket{\psi''}=&\frac{N^{-\frac{n}{2}}}
{\sqrt{n!}}
\sum_{n_1+n_2+\cdots+n_N=n}\\
&\begin{pmatrix}
n\\n_1,n_2,\cdots,n_N
\end{pmatrix}
\prod_{i=1}^N\sqrt{T}r(n_i)A_i^{\dagger n_i}
\ket{0},
\end{aligned}
\end{equation}
where
\begin{equation}
r(n_i)=\left(-\frac{1-T}{T}n_i+1\right)\sqrt{T}^{n_i}.
\end{equation}
The output state, which is the post-selected state after the second $N$-splitter, can be written as
\begin{equation}\label{eq:apb1}
\ket{\psi}_{\text{out}}=
\sqrt{T}^N\frac{a{(N,n)}}{N^n}
\ket{n},
\end{equation}
where
\begin{equation}
\begin{aligned}
&a(N,n)=\sum_{n_1+n_2+\cdots+n_N=n}
\begin{pmatrix}
n\\n_1,n_2,\cdots,n_N
\end{pmatrix}
\prod_{i=1}^N
r(n_i).
\end{aligned}
\end{equation}
After some algebraic manipulations, we find
\begin{equation}\label{eq:apb2}
\begin{aligned}
a(N,n)=\sqrt{T}^n\sum_{k=0}^{N}
\begin{pmatrix}
N\\k
\end{pmatrix}
&
\frac{n!}{(n-N+k)!}N^{n-N+k}\left(\frac{T-1}{T}\right)^{N-k}.
\end{aligned}
\end{equation}
Putting Eq.~(\ref{eq:apb2}) into Eq.~(\ref{eq:apb1}) we arrive at Eq.~(\ref{eq:nlapcn}).

\bibliography{mybib}

\begin{thebibliography}{44}%
\makeatletter
\providecommand \@ifxundefined [1]{%
 \@ifx{#1\undefined}
}%
\providecommand \@ifnum [1]{%
 \ifnum #1\expandafter \@firstoftwo
 \else \expandafter \@secondoftwo
 \fi
}%
\providecommand \@ifx [1]{%
 \ifx #1\expandafter \@firstoftwo
 \else \expandafter \@secondoftwo
 \fi
}%
\providecommand \natexlab [1]{#1}%
\providecommand \enquote  [1]{``#1''}%
\providecommand \bibnamefont  [1]{#1}%
\providecommand \bibfnamefont [1]{#1}%
\providecommand \citenamefont [1]{#1}%
\providecommand \href@noop [0]{\@secondoftwo}%
\providecommand \href [0]{\begingroup \@sanitize@url \@href}%
\providecommand \@href[1]{\@@startlink{#1}\@@href}%
\providecommand \@@href[1]{\endgroup#1\@@endlink}%
\providecommand \@sanitize@url [0]{\catcode `\\12\catcode `\$12\catcode
  `\&12\catcode `\#12\catcode `\^12\catcode `\_12\catcode `\%12\relax}%
\providecommand \@@startlink[1]{}%
\providecommand \@@endlink[0]{}%
\providecommand \url  [0]{\begingroup\@sanitize@url \@url }%
\providecommand \@url [1]{\endgroup\@href {#1}{\urlprefix }}%
\providecommand \urlprefix  [0]{URL }%
\providecommand \Eprint [0]{\href }%
\providecommand \doibase [0]{http://dx.doi.org/}%
\providecommand \selectlanguage [0]{\@gobble}%
\providecommand \bibinfo  [0]{\@secondoftwo}%
\providecommand \bibfield  [0]{\@secondoftwo}%
\providecommand \translation [1]{[#1]}%
\providecommand \BibitemOpen [0]{}%
\providecommand \bibitemStop [0]{}%
\providecommand \bibitemNoStop [0]{.\EOS\space}%
\providecommand \EOS [0]{\spacefactor3000\relax}%
\providecommand \BibitemShut  [1]{\csname bibitem#1\endcsname}%
\let\auto@bib@innerbib\@empty
\bibitem [{\citenamefont {Christ}\ \emph {et~al.}(2012)\citenamefont {Christ},
  \citenamefont {Lupo},\ and\ \citenamefont
  {Silberhorn}}]{christ2012exponentially}%
  \BibitemOpen
  \bibfield  {author} {\bibinfo {author} {\bibfnamefont {A.}~\bibnamefont
  {Christ}}, \bibinfo {author} {\bibfnamefont {C.}~\bibnamefont {Lupo}}, \ and\
  \bibinfo {author} {\bibfnamefont {C.}~\bibnamefont {Silberhorn}},\
  }\href@noop {} {\bibfield  {journal} {\bibinfo  {journal} {New J. Phys.}\
  }\textbf {\bibinfo {volume} {14}},\ \bibinfo {pages} {083007} (\bibinfo
  {year} {2012})}\BibitemShut {NoStop}%
\bibitem [{\citenamefont {Hosseinidehaj}\ and\ \citenamefont
  {Malaney}(2017)}]{hosseinidehaj2017multimode}%
  \BibitemOpen
  \bibfield  {author} {\bibinfo {author} {\bibfnamefont {N.}~\bibnamefont
  {Hosseinidehaj}}\ and\ \bibinfo {author} {\bibfnamefont {R.}~\bibnamefont
  {Malaney}},\ }in\ \href@noop {} {\emph {\bibinfo {booktitle} {2017 IEEE 85th
  Vehicular Technology Conference (VTC Spring)}}}\ (\bibinfo {organization}
  {IEEE},\ \bibinfo {year} {2017})\ pp.\ \bibinfo {pages} {1--5}\BibitemShut
  {NoStop}%
\bibitem [{\citenamefont {Kumar}\ \emph {et~al.}(2019)\citenamefont {Kumar},
  \citenamefont {Tang}, \citenamefont {Wonfor}, \citenamefont {Penty},\ and\
  \citenamefont {White}}]{kumar2019continuous}%
  \BibitemOpen
  \bibfield  {author} {\bibinfo {author} {\bibfnamefont {R.}~\bibnamefont
  {Kumar}}, \bibinfo {author} {\bibfnamefont {X.}~\bibnamefont {Tang}},
  \bibinfo {author} {\bibfnamefont {A.}~\bibnamefont {Wonfor}}, \bibinfo
  {author} {\bibfnamefont {R.}~\bibnamefont {Penty}}, \ and\ \bibinfo {author}
  {\bibfnamefont {I.}~\bibnamefont {White}},\ }\href@noop {} {\bibfield
  {journal} {\bibinfo  {journal} {JOSA B}\ }\textbf {\bibinfo {volume} {36}},\
  \bibinfo {pages} {B109} (\bibinfo {year} {2019})}\BibitemShut {NoStop}%
\bibitem [{\citenamefont {Menicucci}\ \emph {et~al.}(2008)\citenamefont
  {Menicucci}, \citenamefont {Flammia},\ and\ \citenamefont
  {Pfister}}]{menicucci2008one}%
  \BibitemOpen
  \bibfield  {author} {\bibinfo {author} {\bibfnamefont {N.~C.}\ \bibnamefont
  {Menicucci}}, \bibinfo {author} {\bibfnamefont {S.~T.}\ \bibnamefont
  {Flammia}}, \ and\ \bibinfo {author} {\bibfnamefont {O.}~\bibnamefont
  {Pfister}},\ }\href@noop {} {\bibfield  {journal} {\bibinfo  {journal} {Phys.
  Rev. Lett.}\ }\textbf {\bibinfo {volume} {101}},\ \bibinfo {pages} {130501}
  (\bibinfo {year} {2008})}\BibitemShut {NoStop}%
\bibitem [{\citenamefont {Armstrong}\ \emph {et~al.}(2012)\citenamefont
  {Armstrong}, \citenamefont {Morizur}, \citenamefont {Janousek}, \citenamefont
  {Hage}, \citenamefont {Treps}, \citenamefont {Lam},\ and\ \citenamefont
  {Bachor}}]{armstrong2012programmable}%
  \BibitemOpen
  \bibfield  {author} {\bibinfo {author} {\bibfnamefont {S.}~\bibnamefont
  {Armstrong}}, \bibinfo {author} {\bibfnamefont {J.-F.}\ \bibnamefont
  {Morizur}}, \bibinfo {author} {\bibfnamefont {J.}~\bibnamefont {Janousek}},
  \bibinfo {author} {\bibfnamefont {B.}~\bibnamefont {Hage}}, \bibinfo {author}
  {\bibfnamefont {N.}~\bibnamefont {Treps}}, \bibinfo {author} {\bibfnamefont
  {P.~K.}\ \bibnamefont {Lam}}, \ and\ \bibinfo {author} {\bibfnamefont
  {H.-A.}\ \bibnamefont {Bachor}},\ }\href@noop {} {\bibfield  {journal}
  {\bibinfo  {journal} {Nat. Commun.}\ }\textbf {\bibinfo {volume} {3}},\
  \bibinfo {pages} {1} (\bibinfo {year} {2012})}\BibitemShut {NoStop}%
\bibitem [{\citenamefont {Ferrini}\ \emph {et~al.}(2013)\citenamefont
  {Ferrini}, \citenamefont {Gazeau}, \citenamefont {Coudreau}, \citenamefont
  {Fabre},\ and\ \citenamefont {Treps}}]{ferrini2013compact}%
  \BibitemOpen
  \bibfield  {author} {\bibinfo {author} {\bibfnamefont {G.}~\bibnamefont
  {Ferrini}}, \bibinfo {author} {\bibfnamefont {J.-P.}\ \bibnamefont {Gazeau}},
  \bibinfo {author} {\bibfnamefont {T.}~\bibnamefont {Coudreau}}, \bibinfo
  {author} {\bibfnamefont {C.}~\bibnamefont {Fabre}}, \ and\ \bibinfo {author}
  {\bibfnamefont {N.}~\bibnamefont {Treps}},\ }\href@noop {} {\bibfield
  {journal} {\bibinfo  {journal} {New J. Phys.}\ }\textbf {\bibinfo {volume}
  {15}},\ \bibinfo {pages} {093015} (\bibinfo {year} {2013})}\BibitemShut
  {NoStop}%
\bibitem [{\citenamefont {Chen}\ \emph {et~al.}(2014)\citenamefont {Chen},
  \citenamefont {Menicucci},\ and\ \citenamefont
  {Pfister}}]{chen2014experimental}%
  \BibitemOpen
  \bibfield  {author} {\bibinfo {author} {\bibfnamefont {M.}~\bibnamefont
  {Chen}}, \bibinfo {author} {\bibfnamefont {N.~C.}\ \bibnamefont {Menicucci}},
  \ and\ \bibinfo {author} {\bibfnamefont {O.}~\bibnamefont {Pfister}},\
  }\href@noop {} {\bibfield  {journal} {\bibinfo  {journal} {Phys. Rev. Lett.}\
  }\textbf {\bibinfo {volume} {112}},\ \bibinfo {pages} {120505} (\bibinfo
  {year} {2014})}\BibitemShut {NoStop}%
\bibitem [{\citenamefont {Zhuang}\ \emph {et~al.}(2018)\citenamefont {Zhuang},
  \citenamefont {Zhang},\ and\ \citenamefont
  {Shapiro}}]{zhuang2018distributed}%
  \BibitemOpen
  \bibfield  {author} {\bibinfo {author} {\bibfnamefont {Q.}~\bibnamefont
  {Zhuang}}, \bibinfo {author} {\bibfnamefont {Z.}~\bibnamefont {Zhang}}, \
  and\ \bibinfo {author} {\bibfnamefont {J.~H.}\ \bibnamefont {Shapiro}},\
  }\href@noop {} {\bibfield  {journal} {\bibinfo  {journal} {Phys. Rev. A}\
  }\textbf {\bibinfo {volume} {97}},\ \bibinfo {pages} {032329} (\bibinfo
  {year} {2018})}\BibitemShut {NoStop}%
\bibitem [{\citenamefont {Xia}\ \emph {et~al.}(2019)\citenamefont {Xia},
  \citenamefont {Zhuang}, \citenamefont {Clark},\ and\ \citenamefont
  {Zhang}}]{xia2019repeater}%
  \BibitemOpen
  \bibfield  {author} {\bibinfo {author} {\bibfnamefont {Y.}~\bibnamefont
  {Xia}}, \bibinfo {author} {\bibfnamefont {Q.}~\bibnamefont {Zhuang}},
  \bibinfo {author} {\bibfnamefont {W.}~\bibnamefont {Clark}}, \ and\ \bibinfo
  {author} {\bibfnamefont {Z.}~\bibnamefont {Zhang}},\ }\href@noop {}
  {\bibfield  {journal} {\bibinfo  {journal} {Phys. Rev. A}\ }\textbf {\bibinfo
  {volume} {99}},\ \bibinfo {pages} {012328} (\bibinfo {year}
  {2019})}\BibitemShut {NoStop}%
\bibitem [{\citenamefont {Guo}\ \emph {et~al.}(2020)\citenamefont {Guo},
  \citenamefont {Breum}, \citenamefont {Borregaard}, \citenamefont {Izumi},
  \citenamefont {Larsen}, \citenamefont {Gehring}, \citenamefont {Christandl},
  \citenamefont {Neergaard-Nielsen},\ and\ \citenamefont
  {Andersen}}]{guo2020distributed}%
  \BibitemOpen
  \bibfield  {author} {\bibinfo {author} {\bibfnamefont {X.}~\bibnamefont
  {Guo}}, \bibinfo {author} {\bibfnamefont {C.~R.}\ \bibnamefont {Breum}},
  \bibinfo {author} {\bibfnamefont {J.}~\bibnamefont {Borregaard}}, \bibinfo
  {author} {\bibfnamefont {S.}~\bibnamefont {Izumi}}, \bibinfo {author}
  {\bibfnamefont {M.~V.}\ \bibnamefont {Larsen}}, \bibinfo {author}
  {\bibfnamefont {T.}~\bibnamefont {Gehring}}, \bibinfo {author} {\bibfnamefont
  {M.}~\bibnamefont {Christandl}}, \bibinfo {author} {\bibfnamefont {J.~S.}\
  \bibnamefont {Neergaard-Nielsen}}, \ and\ \bibinfo {author} {\bibfnamefont
  {U.~L.}\ \bibnamefont {Andersen}},\ }\href@noop {} {\bibfield  {journal}
  {\bibinfo  {journal} {Nat. Phys.}\ }\textbf {\bibinfo {volume} {16}},\
  \bibinfo {pages} {281} (\bibinfo {year} {2020})}\BibitemShut {NoStop}%
\bibitem [{\citenamefont {Gessner}\ \emph {et~al.}(2020)\citenamefont
  {Gessner}, \citenamefont {Smerzi},\ and\ \citenamefont
  {Pezz{\`e}}}]{gessner2020multiparameter}%
  \BibitemOpen
  \bibfield  {author} {\bibinfo {author} {\bibfnamefont {M.}~\bibnamefont
  {Gessner}}, \bibinfo {author} {\bibfnamefont {A.}~\bibnamefont {Smerzi}}, \
  and\ \bibinfo {author} {\bibfnamefont {L.}~\bibnamefont {Pezz{\`e}}},\
  }\href@noop {} {\bibfield  {journal} {\bibinfo  {journal} {Nat. Commun.}\
  }\textbf {\bibinfo {volume} {11}},\ \bibinfo {pages} {1} (\bibinfo {year}
  {2020})}\BibitemShut {NoStop}%
\bibitem [{\citenamefont {Cai}\ \emph {et~al.}(2017)\citenamefont {Cai},
  \citenamefont {Roslund}, \citenamefont {Ferrini}, \citenamefont {Arzani},
  \citenamefont {Xu}, \citenamefont {Fabre},\ and\ \citenamefont
  {Treps}}]{cai2017multimode}%
  \BibitemOpen
  \bibfield  {author} {\bibinfo {author} {\bibfnamefont {Y.}~\bibnamefont
  {Cai}}, \bibinfo {author} {\bibfnamefont {J.}~\bibnamefont {Roslund}},
  \bibinfo {author} {\bibfnamefont {G.}~\bibnamefont {Ferrini}}, \bibinfo
  {author} {\bibfnamefont {F.}~\bibnamefont {Arzani}}, \bibinfo {author}
  {\bibfnamefont {X.}~\bibnamefont {Xu}}, \bibinfo {author} {\bibfnamefont
  {C.}~\bibnamefont {Fabre}}, \ and\ \bibinfo {author} {\bibfnamefont
  {N.}~\bibnamefont {Treps}},\ }\href@noop {} {\bibfield  {journal} {\bibinfo
  {journal} {Nat. Commun.}\ }\textbf {\bibinfo {volume} {8}},\ \bibinfo {pages}
  {1} (\bibinfo {year} {2017})}\BibitemShut {NoStop}%
\bibitem [{\citenamefont {Fortier}\ and\ \citenamefont
  {Baumann}(2019)}]{fortier201920}%
  \BibitemOpen
  \bibfield  {author} {\bibinfo {author} {\bibfnamefont {T.}~\bibnamefont
  {Fortier}}\ and\ \bibinfo {author} {\bibfnamefont {E.}~\bibnamefont
  {Baumann}},\ }\href@noop {} {\bibfield  {journal} {\bibinfo  {journal}
  {Communications Physics}\ }\textbf {\bibinfo {volume} {2}},\ \bibinfo {pages}
  {1} (\bibinfo {year} {2019})}\BibitemShut {NoStop}%
\bibitem [{\citenamefont {de~Ara{\'u}jo}\ \emph {et~al.}(2014)\citenamefont
  {de~Ara{\'u}jo}, \citenamefont {Roslund}, \citenamefont {Cai}, \citenamefont
  {Ferrini}, \citenamefont {Fabre},\ and\ \citenamefont {Treps}}]{de2014full}%
  \BibitemOpen
  \bibfield  {author} {\bibinfo {author} {\bibfnamefont {R.~M.}\ \bibnamefont
  {de~Ara{\'u}jo}}, \bibinfo {author} {\bibfnamefont {J.}~\bibnamefont
  {Roslund}}, \bibinfo {author} {\bibfnamefont {Y.}~\bibnamefont {Cai}},
  \bibinfo {author} {\bibfnamefont {G.}~\bibnamefont {Ferrini}}, \bibinfo
  {author} {\bibfnamefont {C.}~\bibnamefont {Fabre}}, \ and\ \bibinfo {author}
  {\bibfnamefont {N.}~\bibnamefont {Treps}},\ }\href@noop {} {\bibfield
  {journal} {\bibinfo  {journal} {Phys. Rev. A}\ }\textbf {\bibinfo {volume}
  {89}},\ \bibinfo {pages} {053828} (\bibinfo {year} {2014})}\BibitemShut
  {NoStop}%
\bibitem [{\citenamefont {Roslund}\ \emph {et~al.}(2014)\citenamefont
  {Roslund}, \citenamefont {De~Araujo}, \citenamefont {Jiang}, \citenamefont
  {Fabre},\ and\ \citenamefont {Treps}}]{roslund2014wavelength}%
  \BibitemOpen
  \bibfield  {author} {\bibinfo {author} {\bibfnamefont {J.}~\bibnamefont
  {Roslund}}, \bibinfo {author} {\bibfnamefont {R.~M.}\ \bibnamefont
  {De~Araujo}}, \bibinfo {author} {\bibfnamefont {S.}~\bibnamefont {Jiang}},
  \bibinfo {author} {\bibfnamefont {C.}~\bibnamefont {Fabre}}, \ and\ \bibinfo
  {author} {\bibfnamefont {N.}~\bibnamefont {Treps}},\ }\href@noop {}
  {\bibfield  {journal} {\bibinfo  {journal} {Nat. Photonics}\ }\textbf
  {\bibinfo {volume} {8}},\ \bibinfo {pages} {109} (\bibinfo {year}
  {2014})}\BibitemShut {NoStop}%
\bibitem [{\citenamefont {Gerke}\ \emph {et~al.}(2015)\citenamefont {Gerke},
  \citenamefont {Sperling}, \citenamefont {Vogel}, \citenamefont {Cai},
  \citenamefont {Roslund}, \citenamefont {Treps},\ and\ \citenamefont
  {Fabre}}]{gerke2015full}%
  \BibitemOpen
  \bibfield  {author} {\bibinfo {author} {\bibfnamefont {S.}~\bibnamefont
  {Gerke}}, \bibinfo {author} {\bibfnamefont {J.}~\bibnamefont {Sperling}},
  \bibinfo {author} {\bibfnamefont {W.}~\bibnamefont {Vogel}}, \bibinfo
  {author} {\bibfnamefont {Y.}~\bibnamefont {Cai}}, \bibinfo {author}
  {\bibfnamefont {J.}~\bibnamefont {Roslund}}, \bibinfo {author} {\bibfnamefont
  {N.}~\bibnamefont {Treps}}, \ and\ \bibinfo {author} {\bibfnamefont
  {C.}~\bibnamefont {Fabre}},\ }\href@noop {} {\bibfield  {journal} {\bibinfo
  {journal} {Phys. Rev. Lett.}\ }\textbf {\bibinfo {volume} {114}},\ \bibinfo
  {pages} {050501} (\bibinfo {year} {2015})}\BibitemShut {NoStop}%
\bibitem [{\citenamefont {Caves}(1982)}]{caves1982quantum}%
  \BibitemOpen
  \bibfield  {author} {\bibinfo {author} {\bibfnamefont {C.~M.}\ \bibnamefont
  {Caves}},\ }\href@noop {} {\bibfield  {journal} {\bibinfo  {journal} {Phys.
  Rev. D}\ }\textbf {\bibinfo {volume} {26}},\ \bibinfo {pages} {1817}
  (\bibinfo {year} {1982})}\BibitemShut {NoStop}%
\bibitem [{\citenamefont {Ralph}\ and\ \citenamefont
  {Lund}(2009)}]{ralph2009nondeterministic}%
  \BibitemOpen
  \bibfield  {author} {\bibinfo {author} {\bibfnamefont {T.}~\bibnamefont
  {Ralph}}\ and\ \bibinfo {author} {\bibfnamefont {A.}~\bibnamefont {Lund}},\
  }in\ \href@noop {} {\emph {\bibinfo {booktitle} {AIP Conference
  Proceedings}}},\ Vol.\ \bibinfo {volume} {1110}\ (\bibinfo {organization}
  {American Institute of Physics},\ \bibinfo {year} {2009})\ pp.\ \bibinfo
  {pages} {155--160}\BibitemShut {NoStop}%
\bibitem [{\citenamefont {Zhang}\ and\ \citenamefont
  {Zhang}(2018)}]{zhang2018photon}%
  \BibitemOpen
  \bibfield  {author} {\bibinfo {author} {\bibfnamefont {S.}~\bibnamefont
  {Zhang}}\ and\ \bibinfo {author} {\bibfnamefont {X.}~\bibnamefont {Zhang}},\
  }\href@noop {} {\bibfield  {journal} {\bibinfo  {journal} {Phys. Rev. A}\
  }\textbf {\bibinfo {volume} {97}},\ \bibinfo {pages} {043830} (\bibinfo
  {year} {2018})}\BibitemShut {NoStop}%
\bibitem [{\citenamefont {Pegg}\ \emph {et~al.}(1998)\citenamefont {Pegg},
  \citenamefont {Phillips},\ and\ \citenamefont {Barnett}}]{pegg1998optical}%
  \BibitemOpen
  \bibfield  {author} {\bibinfo {author} {\bibfnamefont {D.~T.}\ \bibnamefont
  {Pegg}}, \bibinfo {author} {\bibfnamefont {L.~S.}\ \bibnamefont {Phillips}},
  \ and\ \bibinfo {author} {\bibfnamefont {S.~M.}\ \bibnamefont {Barnett}},\
  }\href@noop {} {\bibfield  {journal} {\bibinfo  {journal} {Phys. Rev. Lett.}\
  }\textbf {\bibinfo {volume} {81}},\ \bibinfo {pages} {1604} (\bibinfo {year}
  {1998})}\BibitemShut {NoStop}%
\bibitem [{\citenamefont {Mardani}\ \emph {et~al.}(2020)\citenamefont
  {Mardani}, \citenamefont {Shafiei}, \citenamefont {Ghadimi},\ and\
  \citenamefont {Abdi}}]{mardani2020continuous}%
  \BibitemOpen
  \bibfield  {author} {\bibinfo {author} {\bibfnamefont {Y.}~\bibnamefont
  {Mardani}}, \bibinfo {author} {\bibfnamefont {A.}~\bibnamefont {Shafiei}},
  \bibinfo {author} {\bibfnamefont {M.}~\bibnamefont {Ghadimi}}, \ and\
  \bibinfo {author} {\bibfnamefont {M.}~\bibnamefont {Abdi}},\ }\href@noop {}
  {\bibfield  {journal} {\bibinfo  {journal} {Phys. Rev. A}\ }\textbf {\bibinfo
  {volume} {102}},\ \bibinfo {pages} {012407} (\bibinfo {year}
  {2020})}\BibitemShut {NoStop}%
\bibitem [{\citenamefont {Ferreyrol}\ \emph {et~al.}(2010)\citenamefont
  {Ferreyrol}, \citenamefont {Barbieri}, \citenamefont {Blandino},
  \citenamefont {Fossier}, \citenamefont {Tualle-Brouri},\ and\ \citenamefont
  {Grangier}}]{ferreyrol2010implementation}%
  \BibitemOpen
  \bibfield  {author} {\bibinfo {author} {\bibfnamefont {F.}~\bibnamefont
  {Ferreyrol}}, \bibinfo {author} {\bibfnamefont {M.}~\bibnamefont {Barbieri}},
  \bibinfo {author} {\bibfnamefont {R.}~\bibnamefont {Blandino}}, \bibinfo
  {author} {\bibfnamefont {S.}~\bibnamefont {Fossier}}, \bibinfo {author}
  {\bibfnamefont {R.}~\bibnamefont {Tualle-Brouri}}, \ and\ \bibinfo {author}
  {\bibfnamefont {P.}~\bibnamefont {Grangier}},\ }\href@noop {} {\bibfield
  {journal} {\bibinfo  {journal} {Phys. Rev. Lett.}\ }\textbf {\bibinfo
  {volume} {104}},\ \bibinfo {pages} {123603} (\bibinfo {year}
  {2010})}\BibitemShut {NoStop}%
\bibitem [{\citenamefont {Zavatta}\ \emph {et~al.}(2011)\citenamefont
  {Zavatta}, \citenamefont {Fiur{\'a}{\v{s}}ek},\ and\ \citenamefont
  {Bellini}}]{zavatta2011high}%
  \BibitemOpen
  \bibfield  {author} {\bibinfo {author} {\bibfnamefont {A.}~\bibnamefont
  {Zavatta}}, \bibinfo {author} {\bibfnamefont {J.}~\bibnamefont
  {Fiur{\'a}{\v{s}}ek}}, \ and\ \bibinfo {author} {\bibfnamefont
  {M.}~\bibnamefont {Bellini}},\ }\href@noop {} {\bibfield  {journal} {\bibinfo
   {journal} {Nat. Photonics}\ }\textbf {\bibinfo {volume} {5}},\ \bibinfo
  {pages} {52} (\bibinfo {year} {2011})}\BibitemShut {NoStop}%
\bibitem [{\citenamefont {Chrzanowski}\ \emph {et~al.}(2014)\citenamefont
  {Chrzanowski}, \citenamefont {Walk}, \citenamefont {Assad}, \citenamefont
  {Janousek}, \citenamefont {Hosseini}, \citenamefont {Ralph}, \citenamefont
  {Symul},\ and\ \citenamefont {Lam}}]{chrzanowski2014measurement}%
  \BibitemOpen
  \bibfield  {author} {\bibinfo {author} {\bibfnamefont {H.~M.}\ \bibnamefont
  {Chrzanowski}}, \bibinfo {author} {\bibfnamefont {N.}~\bibnamefont {Walk}},
  \bibinfo {author} {\bibfnamefont {S.~M.}\ \bibnamefont {Assad}}, \bibinfo
  {author} {\bibfnamefont {J.}~\bibnamefont {Janousek}}, \bibinfo {author}
  {\bibfnamefont {S.}~\bibnamefont {Hosseini}}, \bibinfo {author}
  {\bibfnamefont {T.~C.}\ \bibnamefont {Ralph}}, \bibinfo {author}
  {\bibfnamefont {T.}~\bibnamefont {Symul}}, \ and\ \bibinfo {author}
  {\bibfnamefont {P.~K.}\ \bibnamefont {Lam}},\ }\href@noop {} {\bibfield
  {journal} {\bibinfo  {journal} {Nat. Photonics}\ }\textbf {\bibinfo {volume}
  {8}},\ \bibinfo {pages} {333} (\bibinfo {year} {2014})}\BibitemShut {NoStop}%
\bibitem [{\citenamefont {Gagatsos}\ \emph {et~al.}(2014)\citenamefont
  {Gagatsos}, \citenamefont {Fiur{\'a}{\v{s}}ek}, \citenamefont {Zavatta},
  \citenamefont {Bellini},\ and\ \citenamefont {Cerf}}]{gagatsos2014heralded}%
  \BibitemOpen
  \bibfield  {author} {\bibinfo {author} {\bibfnamefont {C.}~\bibnamefont
  {Gagatsos}}, \bibinfo {author} {\bibfnamefont {J.}~\bibnamefont
  {Fiur{\'a}{\v{s}}ek}}, \bibinfo {author} {\bibfnamefont {A.}~\bibnamefont
  {Zavatta}}, \bibinfo {author} {\bibfnamefont {M.}~\bibnamefont {Bellini}}, \
  and\ \bibinfo {author} {\bibfnamefont {N.}~\bibnamefont {Cerf}},\ }\href@noop
  {} {\bibfield  {journal} {\bibinfo  {journal} {Phys. Rev. A}\ }\textbf
  {\bibinfo {volume} {89}},\ \bibinfo {pages} {062311} (\bibinfo {year}
  {2014})}\BibitemShut {NoStop}%
\bibitem [{\citenamefont {Ulanov}\ \emph {et~al.}(2015)\citenamefont {Ulanov},
  \citenamefont {Fedorov}, \citenamefont {Pushkina}, \citenamefont {Kurochkin},
  \citenamefont {Ralph},\ and\ \citenamefont {Lvovsky}}]{ulanov2015undoing}%
  \BibitemOpen
  \bibfield  {author} {\bibinfo {author} {\bibfnamefont {A.~E.}\ \bibnamefont
  {Ulanov}}, \bibinfo {author} {\bibfnamefont {I.~A.}\ \bibnamefont {Fedorov}},
  \bibinfo {author} {\bibfnamefont {A.~A.}\ \bibnamefont {Pushkina}}, \bibinfo
  {author} {\bibfnamefont {Y.~V.}\ \bibnamefont {Kurochkin}}, \bibinfo {author}
  {\bibfnamefont {T.~C.}\ \bibnamefont {Ralph}}, \ and\ \bibinfo {author}
  {\bibfnamefont {A.}~\bibnamefont {Lvovsky}},\ }\href@noop {} {\bibfield
  {journal} {\bibinfo  {journal} {Nat. Photonics}\ }\textbf {\bibinfo {volume}
  {9}},\ \bibinfo {pages} {764} (\bibinfo {year} {2015})}\BibitemShut {NoStop}%
\bibitem [{\citenamefont {Haw}\ \emph {et~al.}(2016)\citenamefont {Haw},
  \citenamefont {Zhao}, \citenamefont {Dias}, \citenamefont {Assad},
  \citenamefont {Bradshaw}, \citenamefont {Blandino}, \citenamefont {Symul},
  \citenamefont {Ralph},\ and\ \citenamefont {Lam}}]{haw2016surpassing}%
  \BibitemOpen
  \bibfield  {author} {\bibinfo {author} {\bibfnamefont {J.~Y.}\ \bibnamefont
  {Haw}}, \bibinfo {author} {\bibfnamefont {J.}~\bibnamefont {Zhao}}, \bibinfo
  {author} {\bibfnamefont {J.}~\bibnamefont {Dias}}, \bibinfo {author}
  {\bibfnamefont {S.~M.}\ \bibnamefont {Assad}}, \bibinfo {author}
  {\bibfnamefont {M.}~\bibnamefont {Bradshaw}}, \bibinfo {author}
  {\bibfnamefont {R.}~\bibnamefont {Blandino}}, \bibinfo {author}
  {\bibfnamefont {T.}~\bibnamefont {Symul}}, \bibinfo {author} {\bibfnamefont
  {T.~C.}\ \bibnamefont {Ralph}}, \ and\ \bibinfo {author} {\bibfnamefont
  {P.~K.}\ \bibnamefont {Lam}},\ }\href@noop {} {\bibfield  {journal} {\bibinfo
   {journal} {Nat. Commun.}\ }\textbf {\bibinfo {volume} {7}},\ \bibinfo
  {pages} {1} (\bibinfo {year} {2016})}\BibitemShut {NoStop}%
\bibitem [{\citenamefont {Mi{\v{c}}uda}\ \emph {et~al.}(2012)\citenamefont
  {Mi{\v{c}}uda}, \citenamefont {Straka}, \citenamefont {Mikov{\'a}},
  \citenamefont {Du{\v{s}}ek}, \citenamefont {Cerf}, \citenamefont
  {Fiur{\'a}{\v{s}}ek},\ and\ \citenamefont
  {Je{\v{z}}ek}}]{mivcuda2012noiseless}%
  \BibitemOpen
  \bibfield  {author} {\bibinfo {author} {\bibfnamefont {M.}~\bibnamefont
  {Mi{\v{c}}uda}}, \bibinfo {author} {\bibfnamefont {I.}~\bibnamefont
  {Straka}}, \bibinfo {author} {\bibfnamefont {M.}~\bibnamefont {Mikov{\'a}}},
  \bibinfo {author} {\bibfnamefont {M.}~\bibnamefont {Du{\v{s}}ek}}, \bibinfo
  {author} {\bibfnamefont {N.~J.}\ \bibnamefont {Cerf}}, \bibinfo {author}
  {\bibfnamefont {J.}~\bibnamefont {Fiur{\'a}{\v{s}}ek}}, \ and\ \bibinfo
  {author} {\bibfnamefont {M.}~\bibnamefont {Je{\v{z}}ek}},\ }\href@noop {}
  {\bibfield  {journal} {\bibinfo  {journal} {Phys. Rev. Lett.}\ }\textbf
  {\bibinfo {volume} {109}},\ \bibinfo {pages} {180503} (\bibinfo {year}
  {2012})}\BibitemShut {NoStop}%
\bibitem [{\citenamefont {Kim}\ \emph {et~al.}(2012)\citenamefont {Kim},
  \citenamefont {Lee}, \citenamefont {Ji},\ and\ \citenamefont
  {Nha}}]{kim2012quantum}%
  \BibitemOpen
  \bibfield  {author} {\bibinfo {author} {\bibfnamefont {H.-J.}\ \bibnamefont
  {Kim}}, \bibinfo {author} {\bibfnamefont {S.-Y.}\ \bibnamefont {Lee}},
  \bibinfo {author} {\bibfnamefont {S.-W.}\ \bibnamefont {Ji}}, \ and\ \bibinfo
  {author} {\bibfnamefont {H.}~\bibnamefont {Nha}},\ }\href@noop {} {\bibfield
  {journal} {\bibinfo  {journal} {Phys. Rev. A}\ }\textbf {\bibinfo {volume}
  {85}},\ \bibinfo {pages} {013839} (\bibinfo {year} {2012})}\BibitemShut
  {NoStop}%
\bibitem [{\citenamefont {Yang}\ \emph {et~al.}(2013)\citenamefont {Yang},
  \citenamefont {Zhang}, \citenamefont {Zou}, \citenamefont {Bi},\ and\
  \citenamefont {Lin}}]{yang2013improving}%
  \BibitemOpen
  \bibfield  {author} {\bibinfo {author} {\bibfnamefont {S.}~\bibnamefont
  {Yang}}, \bibinfo {author} {\bibfnamefont {S.}~\bibnamefont {Zhang}},
  \bibinfo {author} {\bibfnamefont {X.}~\bibnamefont {Zou}}, \bibinfo {author}
  {\bibfnamefont {S.}~\bibnamefont {Bi}}, \ and\ \bibinfo {author}
  {\bibfnamefont {X.}~\bibnamefont {Lin}},\ }\href@noop {} {\bibfield
  {journal} {\bibinfo  {journal} {Phys. Rev. A}\ }\textbf {\bibinfo {volume}
  {87}},\ \bibinfo {pages} {024302} (\bibinfo {year} {2013})}\BibitemShut
  {NoStop}%
\bibitem [{\citenamefont {Zhao}\ \emph {et~al.}(2017)\citenamefont {Zhao},
  \citenamefont {Haw}, \citenamefont {Symul}, \citenamefont {Lam},\ and\
  \citenamefont {Assad}}]{zhao2017characterization}%
  \BibitemOpen
  \bibfield  {author} {\bibinfo {author} {\bibfnamefont {J.}~\bibnamefont
  {Zhao}}, \bibinfo {author} {\bibfnamefont {J.~Y.}\ \bibnamefont {Haw}},
  \bibinfo {author} {\bibfnamefont {T.}~\bibnamefont {Symul}}, \bibinfo
  {author} {\bibfnamefont {P.~K.}\ \bibnamefont {Lam}}, \ and\ \bibinfo
  {author} {\bibfnamefont {S.~M.}\ \bibnamefont {Assad}},\ }\href@noop {}
  {\bibfield  {journal} {\bibinfo  {journal} {Phys. Rev. A}\ }\textbf {\bibinfo
  {volume} {96}},\ \bibinfo {pages} {012319} (\bibinfo {year}
  {2017})}\BibitemShut {NoStop}%
\bibitem [{\citenamefont {Hu}\ \emph {et~al.}(2019)\citenamefont {Hu},
  \citenamefont {Al-amri}, \citenamefont {Liao},\ and\ \citenamefont
  {Zubairy}}]{hu2019entanglement}%
  \BibitemOpen
  \bibfield  {author} {\bibinfo {author} {\bibfnamefont {L.}~\bibnamefont
  {Hu}}, \bibinfo {author} {\bibfnamefont {M.}~\bibnamefont {Al-amri}},
  \bibinfo {author} {\bibfnamefont {Z.}~\bibnamefont {Liao}}, \ and\ \bibinfo
  {author} {\bibfnamefont {M.}~\bibnamefont {Zubairy}},\ }\href@noop {}
  {\bibfield  {journal} {\bibinfo  {journal} {Phys. Rev. A}\ }\textbf {\bibinfo
  {volume} {100}},\ \bibinfo {pages} {052322} (\bibinfo {year}
  {2019})}\BibitemShut {NoStop}%
\bibitem [{\citenamefont {Winnel}\ \emph {et~al.}(2020)\citenamefont {Winnel},
  \citenamefont {Hosseinidehaj},\ and\ \citenamefont
  {Ralph}}]{winnel2020generalised}%
  \BibitemOpen
  \bibfield  {author} {\bibinfo {author} {\bibfnamefont {M.}~\bibnamefont
  {Winnel}}, \bibinfo {author} {\bibfnamefont {N.}~\bibnamefont
  {Hosseinidehaj}}, \ and\ \bibinfo {author} {\bibfnamefont {T.~C.}\
  \bibnamefont {Ralph}},\ }\href@noop {} {\bibfield  {journal} {\bibinfo
  {journal} {arXiv:2002.12566}\ } (\bibinfo {year} {2020})}\BibitemShut
  {NoStop}%
\bibitem [{\citenamefont {He}\ \emph {et~al.}(2020)\citenamefont {He},
  \citenamefont {Malaney},\ and\ \citenamefont {Green}}]{he2020global}%
  \BibitemOpen
  \bibfield  {author} {\bibinfo {author} {\bibfnamefont {M.}~\bibnamefont
  {He}}, \bibinfo {author} {\bibfnamefont {R.}~\bibnamefont {Malaney}}, \ and\
  \bibinfo {author} {\bibfnamefont {J.}~\bibnamefont {Green}},\ }\href@noop {}
  {\bibfield  {journal} {\bibinfo  {journal} {IEEE J. Sel. Areas Commun.}\
  }\textbf {\bibinfo {volume} {38}},\ \bibinfo {pages} {528} (\bibinfo {year}
  {2020})}\BibitemShut {NoStop}%
\bibitem [{\citenamefont {Dias}\ and\ \citenamefont
  {Ralph}(2017)}]{dias2017quantum}%
  \BibitemOpen
  \bibfield  {author} {\bibinfo {author} {\bibfnamefont {J.}~\bibnamefont
  {Dias}}\ and\ \bibinfo {author} {\bibfnamefont {T.~C.}\ \bibnamefont
  {Ralph}},\ }\href@noop {} {\bibfield  {journal} {\bibinfo  {journal} {Phys.
  Rev. A}\ }\textbf {\bibinfo {volume} {95}},\ \bibinfo {pages} {022312}
  (\bibinfo {year} {2017})}\BibitemShut {NoStop}%
\bibitem [{\citenamefont {Averchenko}\ \emph {et~al.}(2014)\citenamefont
  {Averchenko}, \citenamefont {Thiel},\ and\ \citenamefont
  {Treps}}]{averchenko2014nonlinear}%
  \BibitemOpen
  \bibfield  {author} {\bibinfo {author} {\bibfnamefont {V.~A.}\ \bibnamefont
  {Averchenko}}, \bibinfo {author} {\bibfnamefont {V.}~\bibnamefont {Thiel}}, \
  and\ \bibinfo {author} {\bibfnamefont {N.}~\bibnamefont {Treps}},\
  }\href@noop {} {\bibfield  {journal} {\bibinfo  {journal} {Phys. Rev. A}\
  }\textbf {\bibinfo {volume} {89}},\ \bibinfo {pages} {063808} (\bibinfo
  {year} {2014})}\BibitemShut {NoStop}%
\bibitem [{\citenamefont {Averchenko}\ \emph {et~al.}(2016)\citenamefont
  {Averchenko}, \citenamefont {Jacquard}, \citenamefont {Thiel}, \citenamefont
  {Fabre},\ and\ \citenamefont {Treps}}]{averchenko2016multimode}%
  \BibitemOpen
  \bibfield  {author} {\bibinfo {author} {\bibfnamefont {V.}~\bibnamefont
  {Averchenko}}, \bibinfo {author} {\bibfnamefont {C.}~\bibnamefont
  {Jacquard}}, \bibinfo {author} {\bibfnamefont {V.}~\bibnamefont {Thiel}},
  \bibinfo {author} {\bibfnamefont {C.}~\bibnamefont {Fabre}}, \ and\ \bibinfo
  {author} {\bibfnamefont {N.}~\bibnamefont {Treps}},\ }\href@noop {}
  {\bibfield  {journal} {\bibinfo  {journal} {New J. Phys.}\ }\textbf {\bibinfo
  {volume} {18}},\ \bibinfo {pages} {083042} (\bibinfo {year}
  {2016})}\BibitemShut {NoStop}%
\bibitem [{\citenamefont {Walschaers}\ \emph {et~al.}(2017)\citenamefont
  {Walschaers}, \citenamefont {Fabre}, \citenamefont {Parigi},\ and\
  \citenamefont {Treps}}]{walschaers2017statistical}%
  \BibitemOpen
  \bibfield  {author} {\bibinfo {author} {\bibfnamefont {M.}~\bibnamefont
  {Walschaers}}, \bibinfo {author} {\bibfnamefont {C.}~\bibnamefont {Fabre}},
  \bibinfo {author} {\bibfnamefont {V.}~\bibnamefont {Parigi}}, \ and\ \bibinfo
  {author} {\bibfnamefont {N.}~\bibnamefont {Treps}},\ }\href@noop {}
  {\bibfield  {journal} {\bibinfo  {journal} {Phys. Rev. A}\ }\textbf {\bibinfo
  {volume} {96}},\ \bibinfo {pages} {053835} (\bibinfo {year}
  {2017})}\BibitemShut {NoStop}%
\bibitem [{\citenamefont {Walschaers}\ \emph {et~al.}(2019)\citenamefont
  {Walschaers}, \citenamefont {Ra},\ and\ \citenamefont
  {Treps}}]{walschaers2019mode}%
  \BibitemOpen
  \bibfield  {author} {\bibinfo {author} {\bibfnamefont {M.}~\bibnamefont
  {Walschaers}}, \bibinfo {author} {\bibfnamefont {Y.-S.}\ \bibnamefont {Ra}},
  \ and\ \bibinfo {author} {\bibfnamefont {N.}~\bibnamefont {Treps}},\
  }\href@noop {} {\bibfield  {journal} {\bibinfo  {journal} {Phys. Rev. A}\
  }\textbf {\bibinfo {volume} {100}},\ \bibinfo {pages} {023828} (\bibinfo
  {year} {2019})}\BibitemShut {NoStop}%
\bibitem [{\citenamefont {Ra}\ \emph {et~al.}(2020)\citenamefont {Ra},
  \citenamefont {Dufour}, \citenamefont {Walschaers}, \citenamefont {Jacquard},
  \citenamefont {Michel}, \citenamefont {Fabre},\ and\ \citenamefont
  {Treps}}]{ra2020non}%
  \BibitemOpen
  \bibfield  {author} {\bibinfo {author} {\bibfnamefont {Y.-S.}\ \bibnamefont
  {Ra}}, \bibinfo {author} {\bibfnamefont {A.}~\bibnamefont {Dufour}}, \bibinfo
  {author} {\bibfnamefont {M.}~\bibnamefont {Walschaers}}, \bibinfo {author}
  {\bibfnamefont {C.}~\bibnamefont {Jacquard}}, \bibinfo {author}
  {\bibfnamefont {T.}~\bibnamefont {Michel}}, \bibinfo {author} {\bibfnamefont
  {C.}~\bibnamefont {Fabre}}, \ and\ \bibinfo {author} {\bibfnamefont
  {N.}~\bibnamefont {Treps}},\ }\href@noop {} {\bibfield  {journal} {\bibinfo
  {journal} {Nat. Phys.}\ }\textbf {\bibinfo {volume} {16}},\ \bibinfo {pages}
  {144} (\bibinfo {year} {2020})}\BibitemShut {NoStop}%
\bibitem [{\citenamefont {Huo}\ \emph {et~al.}(2020)\citenamefont {Huo},
  \citenamefont {Liu}, \citenamefont {Li}, \citenamefont {Cui}, \citenamefont
  {Chen}, \citenamefont {Palivela}, \citenamefont {Xie}, \citenamefont {Li},\
  and\ \citenamefont {Ou}}]{huo2020direct}%
  \BibitemOpen
  \bibfield  {author} {\bibinfo {author} {\bibfnamefont {N.}~\bibnamefont
  {Huo}}, \bibinfo {author} {\bibfnamefont {Y.}~\bibnamefont {Liu}}, \bibinfo
  {author} {\bibfnamefont {J.}~\bibnamefont {Li}}, \bibinfo {author}
  {\bibfnamefont {L.}~\bibnamefont {Cui}}, \bibinfo {author} {\bibfnamefont
  {X.}~\bibnamefont {Chen}}, \bibinfo {author} {\bibfnamefont {R.}~\bibnamefont
  {Palivela}}, \bibinfo {author} {\bibfnamefont {T.}~\bibnamefont {Xie}},
  \bibinfo {author} {\bibfnamefont {X.}~\bibnamefont {Li}}, \ and\ \bibinfo
  {author} {\bibfnamefont {Z.}~\bibnamefont {Ou}},\ }\href@noop {} {\bibfield
  {journal} {\bibinfo  {journal} {Phys. Rev. Lett.}\ }\textbf {\bibinfo
  {volume} {124}},\ \bibinfo {pages} {213603} (\bibinfo {year}
  {2020})}\BibitemShut {NoStop}%
\bibitem [{\citenamefont {Plick}\ \emph {et~al.}(2018)\citenamefont {Plick},
  \citenamefont {Arzani}, \citenamefont {Treps}, \citenamefont {Diamanti},\
  and\ \citenamefont {Markham}}]{plick2018violating}%
  \BibitemOpen
  \bibfield  {author} {\bibinfo {author} {\bibfnamefont {W.~N.}\ \bibnamefont
  {Plick}}, \bibinfo {author} {\bibfnamefont {F.}~\bibnamefont {Arzani}},
  \bibinfo {author} {\bibfnamefont {N.}~\bibnamefont {Treps}}, \bibinfo
  {author} {\bibfnamefont {E.}~\bibnamefont {Diamanti}}, \ and\ \bibinfo
  {author} {\bibfnamefont {D.}~\bibnamefont {Markham}},\ }\href@noop {}
  {\bibfield  {journal} {\bibinfo  {journal} {Physical Review A}\ }\textbf
  {\bibinfo {volume} {98}},\ \bibinfo {pages} {062101} (\bibinfo {year}
  {2018})}\BibitemShut {NoStop}%
\bibitem [{\citenamefont {Cai}\ \emph {et~al.}(2020)\citenamefont {Cai},
  \citenamefont {Roslund}, \citenamefont {Thiel}, \citenamefont {Fabre},\ and\
  \citenamefont {Treps}}]{cai2020quantum}%
  \BibitemOpen
  \bibfield  {author} {\bibinfo {author} {\bibfnamefont {Y.}~\bibnamefont
  {Cai}}, \bibinfo {author} {\bibfnamefont {J.}~\bibnamefont {Roslund}},
  \bibinfo {author} {\bibfnamefont {V.}~\bibnamefont {Thiel}}, \bibinfo
  {author} {\bibfnamefont {C.}~\bibnamefont {Fabre}}, \ and\ \bibinfo {author}
  {\bibfnamefont {N.}~\bibnamefont {Treps}},\ }\href@noop {} {\bibfield
  {journal} {\bibinfo  {journal} {arXiv:2003.05833}\ } (\bibinfo {year}
  {2020})}\BibitemShut {NoStop}%
\bibitem [{\citenamefont {Yang}\ \emph {et~al.}(2012)\citenamefont {Yang},
  \citenamefont {Zhang}, \citenamefont {Zou}, \citenamefont {Bi},\ and\
  \citenamefont {Lin}}]{yang2012continuous}%
  \BibitemOpen
  \bibfield  {author} {\bibinfo {author} {\bibfnamefont {S.}~\bibnamefont
  {Yang}}, \bibinfo {author} {\bibfnamefont {S.}~\bibnamefont {Zhang}},
  \bibinfo {author} {\bibfnamefont {X.}~\bibnamefont {Zou}}, \bibinfo {author}
  {\bibfnamefont {S.}~\bibnamefont {Bi}}, \ and\ \bibinfo {author}
  {\bibfnamefont {X.}~\bibnamefont {Lin}},\ }\href@noop {} {\bibfield
  {journal} {\bibinfo  {journal} {Phys. Rev. A}\ }\textbf {\bibinfo {volume}
  {86}},\ \bibinfo {pages} {062321} (\bibinfo {year} {2012})}\BibitemShut
  {NoStop}%
\end{thebibliography}%

\end{document}